\documentclass[aps,prd,onecolumn,eqsecnum,amsmath,nofootinbib,preprintnumbers]{revtex4}%

\usepackage{etoolbox} 
\usepackage{color,graphicx,float}
\usepackage{amsfonts,amssymb,theorem,mathrsfs,times}
\usepackage{bm}
\usepackage{mathtools}
\usepackage{amsfonts,amssymb,theorem,mathrsfs}
\usepackage{dsfont}
\usepackage{setspace}
\usepackage{ulem}
\usepackage[colorlinks,linkcolor=blue,anchorcolor=blue,citecolor=blue]{hyperref}
\usepackage{cancel}  
\usepackage{booktabs}
\usepackage{multirow}
\usepackage[caption=false,font=footnotesize]{subfig}

{\theorembodyfont{\upshape}
}
{\theorembodyfont{\upshape}
}
{\theorembodyfont{\upshape}
}
{\theorembodyfont{\upshape}
}
{\theorembodyfont{\upshape}
}
{\theorembodyfont{\upshape}
}

\newcommand{\dalm}{\kern1pt\vbox{\hrule height 0.9pt\hbox{\vrule width
0.9pt\hskip 2.5pt\vbox{\vskip 5.5pt}\hskip 3pt\vrule width
0.3pt}\hrule height 0.3pt}\kern1pt}

\begin{document}
\preprint{\hfill {\small{USTC-ICTS/PCFT-25-55}}}

\title{Imaging Signatures of the Israel Junction: Photon Ring Evolution in Dynamical Thin Shell Schwarzschild Spacetimes}

%

\author{ Li-Ming Cao$^{a\, ,b}$
\footnote{e-mail
address: caolm@ustc.edu.cn}
}

\author{ Long-Yue Li$^c$
\footnote{e-mail
address: lilongyue@nbu.edu.cn}
}

\author{ Xia-Yuan Liu$^b$ \footnote{e-mail
address: liuxiayuan@mail.ustc.edu.cn }}


\affiliation{$^a$Peng Huanwu Center for Fundamental Theory, Hefei, Anhui 230026, China}

\affiliation{${}^b$
Interdisciplinary Center for Theoretical Study and Department of Modern Physics,\\
University of Science and Technology of China, Hefei, Anhui 230026,
China}

\affiliation{${}^c$
Institute of Fundamental Physics and Quantum Technology, Department of Physics, School of Physical Science and Technology,
Ningbo University, Ningbo, Zhejiang 315211, China}

\begin{abstract}
    We study the images of black holes by gluing two Schwarzschild spacetimes with a thin shell where the Israel junction conditions are satisfied. 
    By studying the refraction law for null geodesics at the spherical shell, and taking account of the light travel time delay, the images are obtained by ray tracing a geometrically and optically thin accretion disk.
    For a static shell we identify three signatures: a redshift cusp at the shell, a V-shaped profile of the transfer function $r(b)$, and a loss of the one-to-one correspondence between photon spheres and photon rings on the observer's screen.
    During the collapse of the shell, the spacetime evolves from a stage with a single photon sphere inside the shell, through an intermediate stage with double photon spheres, and finally to a spacetime with a single photon sphere outside the shell. 
    However, when the shell is released from a large distance, the corresponding images never show two separate photon rings, even in the stage with two photon spheres. 
    In addition, the motion of the shell leads to a discontinuity in the redshift factor. These signatures provide a practical basis for testing the Israel junction in black hole spacetimes.
\end{abstract}

\maketitle
	
\author{ Xia-Yuan Liu$^b$\footnote{e-mail
		address: liuxiayuan@mail.ustc.edu.cn}}	


\date{\today}
	

\section{Introduction}

In general relativity, joining two spacetimes along a hypersurface is a foundational and important problem. 
Many physical situations naturally exhibit geometry that is defined in pieces:  
models of stellar collapse often describe a fluid interior joined to an exterior Schwarzschild region; 
accretion flows and cosmological phase transitions are likewise idealized as two spacetime regions matched at a common hypersurface, and so on.
Since Darmois’s early systematic treatment of non-singular matchings, it has been clarified that, on a timelike or spacelike matching hypersurface $\Sigma$, the absence of $\delta$-type singularities in the curvature requires continuity of both the induced metric and the extrinsic curvature, namely the Darmois--Lichnerowicz conditions
\cite{MSM_1927__25__1_0,1955trdl.book.....L}. Subsequently, within a distributional framework that allows a $\delta$-source on $\Sigma$, Israel established the correspondence between the jump of the extrinsic curvature and the surface stress--energy tensor, leading to the now standard Israel junction conditions
\cite{Israel:1966rt}:
\begin{equation}
\bigl[K_{ab}\bigr]-h_{ab}\,[K] = -\,8\pi\, S_{ab},
\end{equation}
which interpret geometric discontinuities as a thin shell source localized on the hypersurface; extensions to null hypersurfaces have also been developed
\cite{Barrabes:1991ng}.

Since their introduction, the Israel junction conditions have seen broad applications in gravitational physics.
In the collapse model by Oppenheimer and Snyder,
one glues a pressureless dust interior to an exterior Schwarzschild spacetime along the stellar surface in order to track the boundary evolution and demonstrate horizon formation \cite{Oppenheimer:1939ue}.
In early thin shell wormhole constructions, one ``cuts and pastes'' two Schwarzschild exteriors, localizing matter on the shell and using the junction conditions to relate the surface stress-energy to the throat geometry \cite{Visser:1989kg,Poisson:1995sv}.
In the gravastar scenario, a Schwarzschild exterior is joined to a de Sitter interior by a thin shell. 
From the junction conditions, an effective potential for the shell radius can be derived. The analysis of this potential shows that stable, horizonless compact configurations can exist for suitable equations of state and parameter choices \cite{Mazur:2004fk,Mazur:2001fv,Visser:2003ge}.
Finally, one may glue a flat interior to an exterior Reissner-Nordstr\"om spacetime by means of a charged thin shell to probe weak cosmic censorship
\cite{Boulware:1973tlq}.
Beyond these representative examples, 
the Israel junction conditions have found widespread application across a broad range of contexts in gravitational theory
\cite{Tsukamoto:2021fpp,Capozziello:2024ucm,DeBianchi:2025bgn,Capozziello:2025wwl}.
More generally, thin shell models can serve as useful idealizations in a variety of astrophysical settings. 
For a static shell, they can approximate a thin layer of dark matter around a black hole
\cite{Adhikari:2014lna,Qian:2021qow}. 
For a dynamical collapsing shell, they can describe late-stage fallback of matter during collapse and spherically symmetric accretion
\cite{Liu:2009ts}.

On the observational side, black hole physics is transitioning from pure theoretical study to quantitative tests.
Since the Event Horizon Telescope (EHT) released the first image of M87* in 2019 and of Sgr~A* in 2022, tests at event horizon scales have become possible \cite{EventHorizonTelescope:2019dse,EventHorizonTelescope:2022wkp,Mizuno:2018lxz}.
Within general relativity, the Schwarzschild solution exhibits a distinctive geometric feature: an unstable spherical photon orbit outside the horizon, the photon sphere, which sets the critical boundary of the observed shadow and governs the strong-deflection lensing features and the sequence of higher order photon rings produced by multiple windings of light \cite{Virbhadra:1999nm,Johnson:2019ljv,Vagnozzi:2022moj}.
Thus, studies of photon spheres are key probes of general relativity in the strong field regime \cite{Lupsasca:2024xhq,DiFilippo:2024poc}.

Research on black hole images has a long history. 
Early studies largely concentrated on characterizing the critical curves of the Schwarzschild and Kerr spacetimes \cite{Synge:1966okc,Bardeen:1973tla}, that is, the images governed by the photon sphere.
These curves demarcate bright and dark regions on the observer's screen, and this program has been gradually extended to a variety of gravitational theories.
With improving observational capabilities, theoretical models have begun to systematically include emission from accretion flows, redshift effects, polarization of light, and related processes
\cite{Gralla:2019xty,Hou:2024qqo}. 
As a result, the black hole image is no longer determined solely by the photon sphere, but by the combined effects of spacetime lensing and various astrophysical processes \cite{Hou:2022eev,Meng:2025ivb,Cao:2023par,Cao:2024kht,Li:2025wmd,Uniyal:2025uvc}.
Even so, many existing works still adopt stationary black holes.
In reality, black holes evolve due to collapse, perturbations, or sustained accretion. When imaging such dynamical spacetimes, one must take into account light travel time delays caused by differences in path length. This requirement makes theoretical models closer to actual observations and is becoming increasingly important.
In recent years, studies of imaging in dynamical black hole spacetimes have continued to emerge \cite{zhang2025emergingblackholeshadow,Liang_2025,wang2025dynamicshadowblackhole}.

Motivated by the above considerations, we examine the observable appearance of black holes endowed with a thin spherical shell and aim to identify features that may arise in spacetimes constructed by junctions.
Thin shell spacetimes are concrete realizations of the Israel junction conditions, and by studying their imaging characteristics we can indirectly probe whether Israel junction geometries may be realized in strong gravity environments and place observational constraints on such models.
To the best of our knowledge, most previous imaging studies of spacetimes built from junctions have focused on wormholes \cite{Nedkova:2013msa,Ohgami:2015nra,Peng:2021osd,Macedo:2025ipc}.
Although black holes with shells have been considered, existing works typically model the spherical shell as accreting matter and assess how it modifies otherwise standard static black hole images \cite{Narayan:2019imo,Wu:2024juj}.
In contrast, we investigate how matching distinct Schwarzschild geometries across a spherical shell modifies the spacetime geometry itself, and we analyze the resulting images for both static and dynamical shells.

More concretely, for a static shell we first analyze the image structure and relate those features that differ from standard black hole images to specific physical processes. 
In particular, we explain the influence of light refraction at the shell and the origin of the cusp in the redshift factor, as well as how these effects change the image. 
More importantly, in typical astronomical observations one expects the characteristic photon sphere structure of a spacetime to be faithfully reflected in its photon rings, so that photon ring observations can be used to probe the photon spheres.
The introduction of the shell breaks this correspondence: in a junction spacetime, photon ring observations may no longer faithfully represent the underlying spacetime structure.

Similarly, in the case of a collapsing shell, the photon sphere structure of the spacetime evolves from an inner photon sphere, to a stage with both an inner and outer photon spheres, and finally to an outer photon sphere. One might therefore expect the image to show a corresponding evolution. However, the image on the observer’s screen only goes through a transition from an inner photon ring to an outer photon ring, and a double photon ring structure never appears. 
We explain the absence of a resolvable double ring and identify a finely tuned configuration that exhibits a genuine double photon ring.
In addition, when the shell is moving, the behavior of the redshift factor also changes, from a cusp to a discontinuity.

The paper is organized as follows. Section~\ref{sec:thin-shell} presents the explicit junction of two Schwarzschild spacetimes, derives the shell's equation of motion, and analyzes the dynamics. Section~\ref{sec:shell-dynamics-redshift} discusses light propagation and redshift in the shell spacetime. Section~\ref{sec:StaticImaging} studies images of a static spherical shell. Section~\ref{sec:DynamicImaging} investigates images for a collapsing shell. Section~\ref{Conclusion} summarizes our conclusions and outlook.
Throughout the paper, we work in geometrized units with \(G = c = 1\).

\section{Thin shell junction of two Schwarzschild spacetimes}\label{sec:thin-shell}

In this section, we match two Schwarzschild spacetimes by a spherical shell, derive the equation of motion for the shell, and analyze its dynamics. 
There are already many works on thin shell junctions
\cite{PhysRevD.44.1891,Gao:2008jy}
and on light propagation in such joined spacetimes
\cite{Wang:2020emr}. 
In this and the next section, we mainly give a relatively complete but still brief review of these results in preparation for the later image analysis.

We glue two Schwarzschild spacetimes along a timelike, spherically symmetric thin shell \(\Sigma\) located at \(r=R(\tau)\). 
The metrics on the two sides of the shell are
\begin{equation}
ds_{\pm}^{2}
=-f_{\pm}(r)\,dt_{\pm}^{2}+f_{\pm}(r)^{-1}\,dr^{2}+r^{2}\,d\Omega^{2},
\end{equation}
with
\begin{equation}
    f_{\pm}(r)=1-\frac{2m_{\pm}}{r}, \qquad d\Omega^{2}=d\theta^{2}+\sin^{2}\!\theta\,d\phi^{2}.
\end{equation}
Here the subscript “\(-\)” denotes the interior region \(r<R(\tau)\), while “\(+\)” denotes the exterior region \(r>R(\tau)\).

By the Israel junction conditions \cite{Israel:1966rt}, the induced metric on the matching hypersurface \(\Sigma\) must be continuous, \([h_{ab}]=0\), where $[X]=X_+-X_-$, whereas its normal derivative may be discontinuous.
The resulting jump in the extrinsic curvature, \([K_{ab}]\), encodes a surface stress-energy tensor localized on \(\Sigma\).

We introduce intrinsic coordinates \(\xi^{a}=(\tau,\theta,\phi)\) on \(\Sigma\). The embeddings of the shell into the two bulks are
\begin{equation}
x_{\pm}^{\mu}(\tau,\theta,\phi)=\bigl(t_{\pm}(\tau),\,R(\tau),\,\theta,\,\phi\bigr).
\end{equation}
Denoting \(\dot{t}\equiv dt/d\tau  \) and \( \dot{R} \equiv dR/d\tau\),
the metric on \(\Sigma\) reads
\begin{equation}
\left.ds^{2}\right|_{\Sigma}^{(\pm)}
=
-\Bigl[f_{\pm}(R)\,\dot{t}_{\pm}^{2}-f_{\pm}(R)^{-1}\dot{R}^{2}\Bigr]\,d\tau^{2}
+R^{2}d\Omega^{2}.
\end{equation}
Choosing \(\tau\) as the shell’s proper time, the normalization of the four velocity gives
\begin{equation}
    \dot{t}_{\pm}=\frac{\beta_{\pm}}{f_{\pm}(R)},
    \qquad
    \beta_{\pm}\equiv\sqrt{\dot{R}^{2}+f_{\pm}(R)}.
    \label{eq:t-dot}
\end{equation}
Let \(U_{\pm}^{\mu}=(\dot{t}_{\pm},\dot{R},0,0)\) be the shell’s four velocity, and \(n_{\pm}^{\mu}\) the unit normal vectors on the two sides. They satisfy
\(g^{(\pm)}_{\mu\nu}U_{\pm}^{\mu}n_{\pm}^{\nu}=0\) and \(g^{(\pm)}_{\mu\nu}n_{\pm}^{\mu}n_{\pm}^{\nu}=+1\).
We choose \(n^{r}_{\pm}>0\), i.e. the normals point toward increasing \(r\), which yields
\begin{equation}
n_{\pm}^{\mu}=\left(\frac{\dot{R}}{f_{\pm}(R)},\,\beta_{\pm},\,0,\,0\right).
\label{eq:normal}
\end{equation}

So far we have only required continuity of the induced metric along the shell. The expressions in Eq.~\eqref{eq:normal} are computed separately in the two bulks and, a priori, need not represent the same geometric vector on \(\Sigma\). 
However, if the unit normals from the two sides are not identified as a single normal field on \(\Sigma\), the jump \([K_{ab}]\) and the associated distributional definitions of the connection and curvature become ambiguous. This issue is more pronounced when three or more bulk spacetimes are glued along a single hypersurface. See Ref.~\cite{Shen:2024dun} and Appendix~A for a detailed discussion.
To ensure a well-defined differentiable structure and distributional curvature at the shell, we therefore identify the unit normals on the two sides at \(\Sigma\):
\begin{equation}
n\big|_{\Sigma}\equiv n_{+}\big|_{\Sigma}=n_{-}\big|_{\Sigma}.
\end{equation}
Accordingly, when matching along a thin shell using the Israel conditions, we impose only continuity of the induced metric across \(\Sigma\), and we implicitly identify the unit normals on the two sides. Thus the spacetime metric is continuous across \(\Sigma\).


Besides the Schwarzschild coordinates on each side, it is convenient to introduce Gaussian normal coordinates (GNC) in a neighborhood of the shell. Starting from the unit normal \(n^{\mu}\) on \(\Sigma\), shoot geodesics orthogonal to the shell and use their proper length as a coordinate \(\ell\) (with \(\ell=0\) on \(\Sigma\) and positive toward the “\(+\)” side). Then we can obtain the metric in GNC:
\begin{equation}
ds^{2}=d\ell^{2}+h_{ij}(\ell,\xi)\,d\xi^{i}d\xi^{j},
\end{equation}
where \(h_{ij}(0, \xi)\) is the induced metric on \(\Sigma\).
It is worth emphasizing that, even if the metric is continuous, the Schwarzschild coordinate bases \(\partial_{t_{\pm}}, \partial_{r_{\pm}}\) generally do not coincide on \(\Sigma\). This is precisely why the conserved quantities of a photon typically change as it crosses the shell. We now derive the corresponding transformations.

On \(\Sigma\) there are three natural pairs of basis vectors: the bulk coordinate bases \((\partial_{t_{+}},\partial_{r_{+}})\), \((\partial_{t_{-}},\partial_{r_{-}})\), and the GNC basis \((\partial_{\tau},\partial_{\ell})\). 
The angular coordinate basis vectors coincide on \(\Sigma\) because we identify the angular coordinates \((\theta,\phi)\) on the two sphere \(S^2\).
Using the shell four velocity \(U_{\pm}^{\mu}\) and the unit normal \(n_{\pm}^{\mu}\),
\begin{equation}
\begin{pmatrix}\partial_{\tau}\\[2pt]\partial_{\ell}\end{pmatrix}
=
\begin{pmatrix}
\dot{t}_{\pm} & \dot{R}\\
\dot{R}/f_{\pm} & \beta_{\pm}
\end{pmatrix}
\begin{pmatrix}\partial_{t_{\pm}}\\[2pt]\partial_{r_{\pm}}\end{pmatrix}.
\label{eq:Apm}
\end{equation}
We denote the transformation matrix by \(A_{\pm}\).
From Eq.~\eqref{eq:Apm} we obtain the transformation between the bulk coordinate bases:
\begin{equation}
\begin{pmatrix}\partial_{t_+}\\[2pt]\partial_{r_+}\end{pmatrix}
=A_{+}^{-1}A_{-}
\begin{pmatrix}\partial_{t_-}\\[2pt]\partial_{r_-}\end{pmatrix}
=
\begin{pmatrix}
\dfrac{\beta_{+}\beta_{-}-\dot{R}^{2}}{f_{-}} & \dot{R}(\beta_{+}-\beta_{-})\\[8pt]
\dfrac{\dot{R}(\beta_{+}-\beta_{-})}{f_{+}f_{-}} & \dfrac{\beta_{+}\beta_{-}-\dot{R}^{2}}{f_{+}}
\end{pmatrix}
\begin{pmatrix}\partial_{t_-}\\[2pt]\partial_{r_-}\end{pmatrix}.
\label{eq:M-matrix}
\end{equation}
If \(m_{+}=m_{-}\), this matrix is the identity. In the static case,
\begin{equation}
\partial_{t_{+}}=\sqrt{\frac{f_{+}(R)}{f_{-}(R)}}\,\partial_{t_{-}},
\qquad
\partial_{r_{+}}=\sqrt{\frac{f_{-}(R)}{f_{+}(R)}}\,\partial_{r_{-}} \,.
\label{eq:StaticTrans}
\end{equation}
Thus, although \(r_{+}=r_{-}=R\) on \(\Sigma\), the radial basis vectors \(\partial_{r_{\pm}}\) need not coincide there.

Having established the geometric junction, we now turn to the shell dynamics. For a timelike, spherically symmetric thin shell at \(r=R(\tau)\), the extrinsic curvatures on the two sides are
\begin{equation}
\left.K^{\tau}{}_{\tau}\right|_{\pm}
=\frac{\ddot{R}+\tfrac{1}{2}f'_{\pm}(R)}{\beta_{\pm}},
\qquad
    \left.K^{\theta}{}_{\theta}\right|_{\pm}
=\left.K^{\phi}{}_{\phi}\right|_{\pm}
=\frac{\beta_{\pm}}{R}.
\label{eq:Ktheta}
\end{equation}
We consider a pressureless thin shell with surface stress-energy tensor
\(S^{i}{}_{j}=\mathrm{diag}(-\sigma,0,0)\).
From the Israel conditions \cite{Israel:1966rt}
\begin{equation}
\bigl[K^{i}{}_{j}\bigr]-\delta^{i}{}_{j}\,[K]=-8\pi S^{i}{}_{j},
\end{equation}
we obtains the equation of motion
\begin{equation}
\beta_+ - \beta_-
=-4\pi\sigma R.
\label{eq:basic}
\end{equation}
The conservation equation is
\begin{equation}
\dot{\sigma}+2\sigma\,\frac{\dot{R}}{R}=0.
\label{eq:surf-cons}
\end{equation}
Requiring \(\sigma \ge 0\) (nonnegative surface energy density) and invoking Eq.~\eqref{eq:basic} yields \(m_- \le m_+\).
We define the shell mass by
\begin{equation}
    m_{\mathrm{sh}} \equiv 4\pi R^{2}\sigma(\tau).
\end{equation}
Eq.~\eqref{eq:surf-cons} shows that \(m_{\mathrm{sh}}\) is constant. 
For a dust shell with surface density $\sigma$, the conserved rest
mass $m_{\rm sh}$ sets the amount of matter localized on the junction.
The mass difference $\Delta m \equiv m_{+}-m_{-}$ represents the total energy carried by
the shell as seen from infinity, including kinetic and gravitational binding contributions.
Using Eq.~\eqref{eq:basic} with \(m_{\mathrm{sh}}=\mathrm{const}\) yields
\cite{Wang:2020emr}
\begin{equation}
\dot R^{\,2} =
\frac{R^{2}}{4m_{\mathrm{sh}}^{2}}([f])^{2}
+\frac{m_{\mathrm{sh}}^{2}}{4R^{2}}
-\frac{f_{+}+f_{-}}{2}.
\label{eq:D-def}
\end{equation}
Furthermore, combining this with \( \dot{t}_\pm \) Eq.~\eqref{eq:t-dot}, we obtain 
\begin{equation}
\frac{dR}{dt_{+}}
=\frac{\dot{R}}{\dot{t}_{+}}
=\pm\,\frac{f_{+}(R)\,|\dot R|}{\beta_{+}(R)},
\qquad
\frac{dR}{dt_{-}}
=\frac{\dot{R}}{\dot{t}_{-}}
=\pm\,\frac{f_{-}(R)\,|\dot R|}{\beta_{-}(R)},
\label{eq:dRdt}
\end{equation}
where the plus sign denotes outward motion of the shell, and the minus sign denotes inward collapse.

\begin{figure}[htbp]
    \centering
        \includegraphics[width= 0.5 \linewidth]{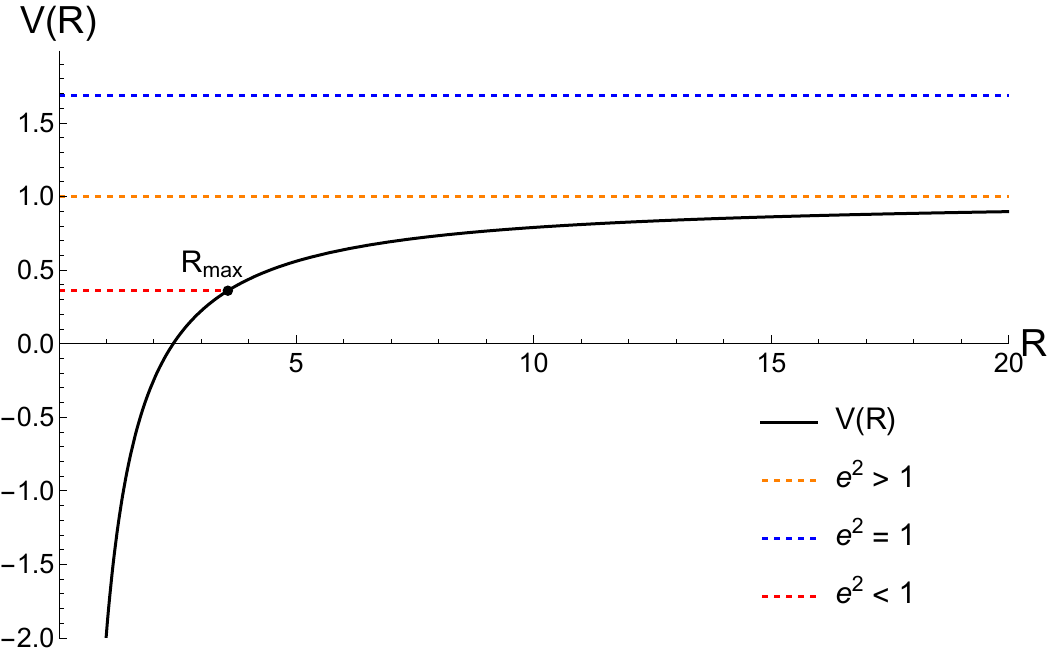}
    \caption{Schematic effective potential $V(R)$ for the thin spherical shell. The solid black curve shows $V(R)$; the horizontal dashed lines in red, blue, and orange mark $e^{2}$ corresponding to the cases $e^{2}\!<\!1$, $e^{2}\!=\!1$, and $e^{2}\!>\!1$, respectively. For $e^{2}<1$, the shell undergoes bound motion: it can reach at most the turning point $R_{\max}$, i.e., the intersection of the red dashed line with the black curve $V(R)$.
}
    \label{VRSchmaticFigure}
\end{figure}

Specializing Eq.~\eqref{eq:D-def} to the Schwarzschild form, we obtain
\begin{equation}
\dot R^{\,2}=e^{2}-V(R),
\label{eq:Rdot-simple}
\end{equation}
with \(e\equiv (m_{+}-m_{-})/m_{\mathrm{sh}}\) and
\begin{equation}
    V(R)=1-\frac{m_{+}+m_{-}}{R}-\frac{m_{\mathrm{sh}}^{\,2}}{4R^{2}}\,.
\end{equation}
Requiring \(\sigma\ge 0\) implies \(e\ge 0\). For \(R>0\) the effective potential is monotonically increasing and concave, i.e.\ \(V'(R)>0\) and \(V''(R)<0\). Fig.~\ref{VRSchmaticFigure} schematically illustrates \(V(R)\). When \(e>1\) the motion is unbound and the shell can expand to spatial infinity; when \(e=1\) the motion is critical and the shell reaches infinity with vanishing asymptotic velocity; when \(e<1\) there is a turning point \(R_{\max}\) defined by \(V(R_{\max})=e^{2}\). 
Consequently, a pressureless shell admits no static, stable configuration, and static stability therefore requires nonzero pressure. Furthermore, any static shell must satisfy the necessary geometric constraint \(R \ge 3m_{-}\), i.e., it must lie outside the interior photon sphere~\cite{PhysRevD.44.1891}.

\section{Null geodesics in the joined spacetime}\label{sec:shell-dynamics-redshift}

Although the global metric of the joined spacetime is continuous, 
the Schwarzschild coordinate bases on the two sides of the shell are different. 
Consequently, a photon’s conserved quantities and trajectory change as it crosses the shell, since the Killing vectors inside and outside are different. This directly impacts the subsequent imaging analysis.
Since we work with spherically symmetric spacetimes, we henceforth restrict to the equatorial plane, $\theta=\pi/2$.

Within a given bulk region, a photon follows the null geodesics of that region. When it passes through the shell, its trajectory is refracted, as if the ray were transmitted from one medium to another. After crossing the shell, the motion is again governed by the null geodesic equation in the new bulk region.
The key task is therefore to determine the precise transmission law at the shell.

From the results above, the metric is continuous at the shell, whereas its normal derivative is not. Hence the Levi–Civita connection \( \Gamma^{\mu}{}_{\alpha\beta} \) is stepwise discontinuous across \(\Sigma\), without any \(\delta\)-type singularity; the \(\delta\)-term appears in the curvature \(R^{\rho}{}_{\sigma\mu\nu}\sim \partial_{\ell}\Gamma\) and is proportional to the jump \([K_{ab}]\). 
For a null geodesic with tangent \(k^{\mu} \equiv dx^{\mu}/d\lambda\), 
integrating the null geodesic equation across a small neighborhood of the crossing parameter \(\lambda_{0}\) in a common coordinate chart across the shell gives
\begin{equation}
\Delta k^{\mu}
= -\!\int_{\lambda_{0}-\epsilon}^{\lambda_{0}+\epsilon}\Gamma^{\mu}{}_{\alpha\beta}k^{\alpha}k^{\beta}\,d\lambda
\longrightarrow 0 \;\;(\epsilon\to 0),
\label{eq:k-cont-restate}
\end{equation}
which implies that the geodesic tangent vector is continuous across $\Sigma$ as a geometric vector. However, because the coordinates and their bases differ on both side of the shell in Eq.~\eqref{eq:M-matrix}, its components are generally discontinuous:
\begin{equation}
\begin{pmatrix}k_{t_+}\\[2pt]k_{r_+}\end{pmatrix}
=
\begin{pmatrix}
\dfrac{\beta_{+}\beta_{-}-\dot{R}^{2}}{f_{-}} & \dot{R}(\beta_{+}-\beta_{-})\\[8pt]
\dfrac{\dot{R}(\beta_{+}-\beta_{-})}{f_{+}f_{-}} & \dfrac{\beta_{+}\beta_{-}-\dot{R}^{2}}{f_{+}}
\end{pmatrix}
\begin{pmatrix}k_{t_-}\\[2pt]k_{r_-}\end{pmatrix},
\qquad
k_{\phi_+}=k_{\phi_-}\equiv L ,
\label{eq:kTrans-restate}
\end{equation}
so the energy \(E=-k_t\) changes at the shell, whereas the angular momentum \(L=k_{\phi}\) is preserved. 
From Eq.~\eqref{eq:kTrans-restate}, the energy transforms as
\begin{equation}
E_{+}
=
\frac{\beta_{+}\beta_{-}-\dot{R}^{2}}{f_{-}(R)}\,E_{-}
-
\dot{R}\,(\beta_{+}-\beta_{-})\;
s\;\frac{1}{f_{-}(R)}
\sqrt{\,E_{-}^{2}-\frac{f_{-}(R)\,L^{2}}{R^{2}}\,},
\label{eq:EnergyTrans}
\end{equation}
where \(s=\mathrm{sgn}(k^{r}_{\pm})\).
In the reconstruction of black hole images, 
it is often convenient to use the impact parameter \(b_{\pm}\equiv L/E_{\pm}\), 
whose transformation is
\begin{equation}
b_{+}=\frac{L}{E_{+}}
=\frac{b_{-}}{\,\frac{\beta_{+}\beta_{-}-\dot{R}^{2}}{f_{-}(R)}\,-s\,\dot{R}\,(\beta_{+}-\beta_{-})\,f_{-}(R)^{-1}\!
\sqrt{\,1-\dfrac{f_{-}(R)}{R^{2}}\,b_{-}^{2}\,}\,}.
\label{eq:bTrans}
\end{equation}

With these relations in hand, we now discuss ``refraction'' at the shell.
Its effect will appear explicitly in ray tracing plots and ultimately in the images.
Consider the photon’s trajectory in the equatorial plane \((x,y)=(r\cos\phi,\; r\sin\phi)\). The angle \(\alpha\) between the photon’s spatial velocity and the radial direction in the equatorial plane satisfies
\begin{equation}
\sin \alpha \equiv \frac{r k^{\phi}}{\sqrt{\left(k^r\right)^2 + \left(r k^{\phi}\right)^2}},
\end{equation}
where we define \(\alpha\) as the angle with respect to the radial direction, \(0\le \alpha\le \pi/2\). From Eq.~\eqref{eq:kTrans-restate} we obtain the relation between the “refracted’’ angles \(\alpha_+\) and \(\alpha_-\):
\begin{equation}
\sin\alpha_{+}=
\left[1+\frac{\left(
s\bigl(\beta_{+}(R)\beta_{-}(R)-\dot{R}^{2}\bigr)\sqrt{\,1-\sin^{2}\alpha_{-}\,}
-\dot{R}\,(\beta_{+}(R)-\beta_{-}(R))\sqrt{\,1-(1-f_{-}(R))\sin^{2}\alpha_{-}\,}
\right)^{2}}{f_{-}(R)^{2}\sin^{2}\alpha_{-}}
\right]^{-1/2}
\end{equation}

For a static shell, the transformations of the energy, impact parameter, and refraction angle simplify to
\begin{equation}
E_{+}=\sqrt{\frac{f_{+}(R)}{f_{-}(R)}}\,E_{-},
\qquad
b_{+}=\sqrt{\frac{f_{-}(R)}{f_{+}(R)}}\, b_{-},
\qquad
\frac{\sin \alpha_+}{\sin \alpha_-}=\sqrt{\frac{1+\frac{f_-(R)}{f_+(R)}\left(\frac{R^2}{b_+^2}-f_+(R)\right)}{1+\left(\frac{R^2}{b_+^2}-f_+(R)\right)}}\,.
\label{eq:Statictrans}
\end{equation}
From the null constraint,
we have $\frac{R^2}{b_+^2}-f_+(R) \geq 0$.
Thus, for a static shell, a photon exiting the shell is redshifted and its “refraction angle" increases.

We now clarify what we mean by the “refraction angle”.
We do not refer to the incident angle measured in an orthonormal frame comoving with the shell. 
Since the photon four momentum is continuous as a geometric vector, the direction measured in that local frame does not jump across the shell. 
Instead, our “refraction angle’’ is defined with respect to the static observers in the inner and outer Schwarzschild regions. 
Owing to the difference in the Killing fields across the shell, the static observers on either side are defined differently.  
As Eq.~\eqref{eq:M-matrix} shows, their orthonormal bases do not match at the shell. Consequently, observer-dependent quantities such as the incident and outgoing angles, as well as the frequency, undergo a change across the shell.
From Eq.~\eqref{eq:Statictrans}, we see that the incident and outgoing angles are generally different across the shell.
To streamline notation and make contact with the optics analogy, we
introduce $n_\pm$ through the Snell-like relation
\begin{equation} 
    \frac{n_-}{n_+}=\frac{\sin\alpha_+}{\sin\alpha_-}.
\end{equation}
Here we can view $n_\pm$ as effective refractive indices on the inner and outer sides of the shell.
Requiring a nonnegative surface energy density implies $\sin\alpha_+ \geq \sin\alpha_-$, and hence $n_- \geq n_+$.
The effect of refraction on the image will be clarified in the section dedicated to imaging.
We note, however, that refractive index interpretations in gravitational settings have been discussed in the literature
\cite{Ramezani-Aval:2024gwv,Masghatian:2025vtr,Nouri-Zonoz:2022vrj}, 
but there is still no unique, universally accepted definition.
In our work, we do not attempt to define a refractive index in a fundamental sense. Instead, we introduce this notion simply as a convenient way to describe how the angle between the photon propagation direction and the radial direction changes when the light ray crosses the shell. Since our specific setting and the aspects we emphasize differ from those in the above studies, the resulting relations and their explicit forms may accordingly be different.

The changes in photon energy when light crosses the shell are encoded in the redshift factor, which directly influences the final image.
For collapsing shells, see Ref.~\cite{cbzn-pzrw} for a systematic treatment; here we briefly summarize and adapt the relevant results.
Define \(g=\omega_{\text{obs}}/\omega_{\text{em}}\). In a static, spherically symmetric background, frequencies are defined with respect to static observers, then
\begin{equation}
g\equiv \frac{\omega_{\mathrm{obs}}}{\omega_{\mathrm{em}}}
= \frac{\sqrt{f_{\mathrm{em}}(r_{\mathrm{em}})}}{\sqrt{f_{+}(r_{\mathrm{obs}})}}
\frac{E_{+}}{E_{\mathrm{em}}}.
\label{eq:g-def}
\end{equation}
Here \(f_{\text{em}}(r)\) and \(E_{\text{em}}\) denote the metric function and the photon energy at the emission point, respectively. 
Each time the light crosses the shell, its energy is transformed according to Eq.~\eqref{eq:EnergyTrans}. 
Hence the total redshift factor is the product of the energy change factors accumulated over all crossings:
\begin{equation}
g=\frac{\sqrt{f_{\rm em}(r_{\rm em})}}{\sqrt{f_{+}(r_{\rm obs})}}
\prod_{j}\!\left(\frac{E_{\text{to}}}{E_{\text{from}}}\right)_{R_j},
\end{equation}
where \(R_j\) is the shell radius at the \(j\)-th crossing. 
The quantities \(E_{\text{to}}\) and \(E_{\text{from}}\) are the photon energies just after and just before that crossing, as measured by static observers on either side of the shell.

When the shell is static, Eq.~\eqref{eq:g-def} simplifies so that shell motion makes no contribution and the redshift depends solely on the emission location.
If the photon is emitted inside the shell,
\begin{equation}
    g
    = \frac{\sqrt{f_{-}(r_{\mathrm{em}})}}{\sqrt{f_{+}(r_{\mathrm{obs}})}}
      \sqrt{\frac{f_+(R)}{f_-(R)}} .
\label{eq:Staticgfactor}
\end{equation}
If the photon is emitted outside the shell, the redshift factor has the standard form \(g=\sqrt{f_{+}(r_{\mathrm{em}})/f_{+}(r_{\mathrm{obs}})}\). As the emission radius passes through the shell radius \(R_{\rm sh}\) (the radial coordinate of the shell), the redshift is continuous but not differentiable at the shell.
For a moving shell, the redshift factor can even be discontinuous because the four velocity of static observers \(\partial_{t}/\sqrt{f}\) changes across the interface. When the shell is at rest this field is continuous, as in Eq.~\eqref{eq:StaticTrans}. With shell motion, Eq.~\eqref{eq:M-matrix} shows that it becomes discontinuous at the shell.
The corresponding effects will be illustrated below.

\section{Static Shell Imaging}\label{sec:StaticImaging}

Having completed the junction of two Schwarzschild spacetimes and the analysis of null geodesics, we now turn to the imaging features of the resulting joined geometry. We begin with a static spherical shell for two reasons. First, the static case is simpler and allows us to identify the concrete impact of the shell on the image and the underlying mechanism, thereby laying the groundwork for understanding the various effects present for a dynamical shell. 
Second, the geometry obtained by matching two spacetimes along a static thin shell is of practical interest in its own right. For example, 
as mentioned in the Introduction,
static shells are sometimes employed as idealized models of dark matter distributions around black holes
\cite{Adhikari:2014lna,Qian:2021qow}.

Varying the shell location substantially reshapes the global geometry. 
For the resulting image, the most important change is the altered photon sphere configuration.
In Schwarzschild spacetime, the photon sphere at \(r=3 M\) consists of unstable circular null geodesics. Its observational feature is the bright ``photon ring'' on the observer’s screen. Observations of this ring have become an important probe of general relativity in the strong field regime. 
Therefore, the photon sphere is not only a highly sensitive probe but also a distinctive structural characteristic of the spacetime geometry.
In what follows, we focus on the photon sphere and use image features to deepen our understanding of the geometry of the junction spacetime.

\begin{figure}[htbp]
    \centering
        \includegraphics[width= 0.5 \linewidth]{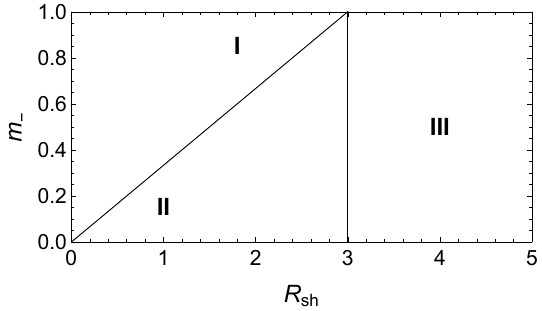}
    \caption{Photon sphere structure as a function of the inner Schwarzschild mass \(m_{-}\) and the shell radius \(R\). 
    We choose $m_{+}=1$.
    Region I contains only the outer photon sphere at \(r_{\rm ph}^{+}=3m_{+}\); 
    Region II contains two photon spheres at \(r_{\rm ph}^{+}=3m_{+}\) and \(r_{\rm ph}^{-}=3m_{-}\); 
    Region III contains only the inner photon sphere at \(r_{\rm ph}^{-}=3m_{-}\). 
    }
    \label{PhaseDiagramStatic}
\end{figure}

Fig.~\ref{PhaseDiagramStatic} shows how the photon sphere structure in the junction spacetime changes as the inner Schwarzschild mass \(m_{-}\) and the shell radius \(R_{\text{sh}}\) vary. 
For ease of comparison with a standard Schwarzschild black hole and to simplify the discussion, we fix \(m_{+}=1\). The parameter space is uniquely characterized by \((m_{-},\,R_{\text{sh}})\). 
As the shell moves inward from larger to smaller radii (from region~III to region~II and finally to region~I), the geometry changes from one with only the inner photon sphere at \(r=3m_{-}\), to a double photon sphere structure with both inner and outer photon spheres, and ultimately to one with only the outer photon sphere at \(r=3m_{+}\).
Here we consider a static shell and move its position by hand, while in the next section we will study a collapsing shell, where the photon sphere structure changes dynamically during the collapse. 
These two situations are different.

To obtain the black hole images, it is essential to specify the accretion model. In what follows we adopt a geometrically thin, optically thin equatorial disk extending from the inner Schwarzschild event horizon to infinity:
\begin{equation}
    I_{\text{em}}(r) =
    \begin{cases}
      \dfrac{1}{\bigl(r - r_{h_-} + 1\bigr)^3}, & \text{if } r \ge r_{h_-}, \\[6pt]
      0, & \text{if } r < r_{h_-}.
    \end{cases}
    \label{accretiondisk}
\end{equation}
where \(r_{h_-}\) is the event horizon radius of the inner Schwarzschild region. 
We adopt this model precisely because our interest lies in image features controlled by the photon sphere. For a geometrically thin and optically thin disk, the photon ring of the final image is largely insensitive to the detailed radial emissivity, and extending the disk to the inner Schwarzschild horizon ensures that all trajectories relevant to the ring are included.
For further discussion and comparisons of accretion disk models, see our previous work~\cite{Cao:2023par, Cao:2024kht, Li:2025wmd}.
Accounting for the redshift factor, 
the observed specific intensity reads
\begin{equation}
    I_{\text{obs}}
    = \sum_{n} g_n^{4}\, I_{\text{em}}(r)\bigl|_{r=r_{n}},
\end{equation}
where \(g \equiv \omega_{\text{obs}}/\omega_{\text{em}}\) is the redshift factor, and \(r_{n}\) is the emission radius at the \(n\)th intersection of the light ray with the accretion disk. 
We place the observer at \((t_{o},\, r_{o}=50,\, \phi_{o}=0)\), restrict to the equatorial plane \((\theta=\pi/2)\), and fix \(m_{+}=1\).

\begin{figure}[htbp]
    \centering
        \includegraphics[width= 1 \linewidth]{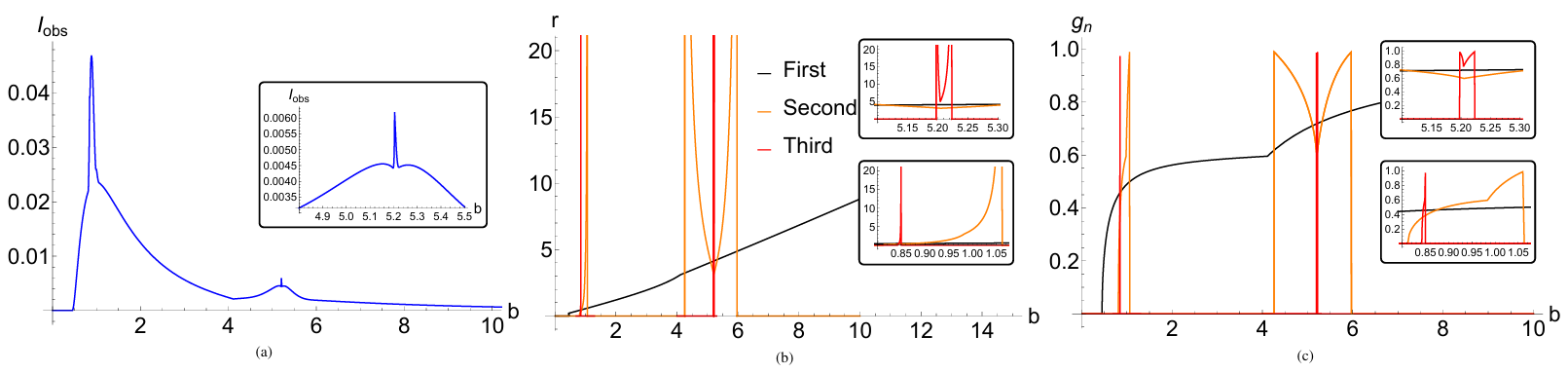}
    \caption{For parameters \(m_-=0.1\) and \(R_{\rm sh}=3.1\). (a) Observed intensity \(I_{\rm obs}\) versus impact parameter \(b\). (b) Transfer function \(r(b)\) for rays with different numbers of disk crossings. (c) Corresponding redshift factors \(g_{n}(b)\). In panels (b) and (c), black, orange, and red curves denote rays that intersect the disk once, twice, and three times, respectively.
    }
    \label{mminus0.1,mplus1,ms3.1Refractioneffect}
\end{figure}

\begin{figure}[htbp]
    \centering
        \includegraphics[width= 1 \linewidth]{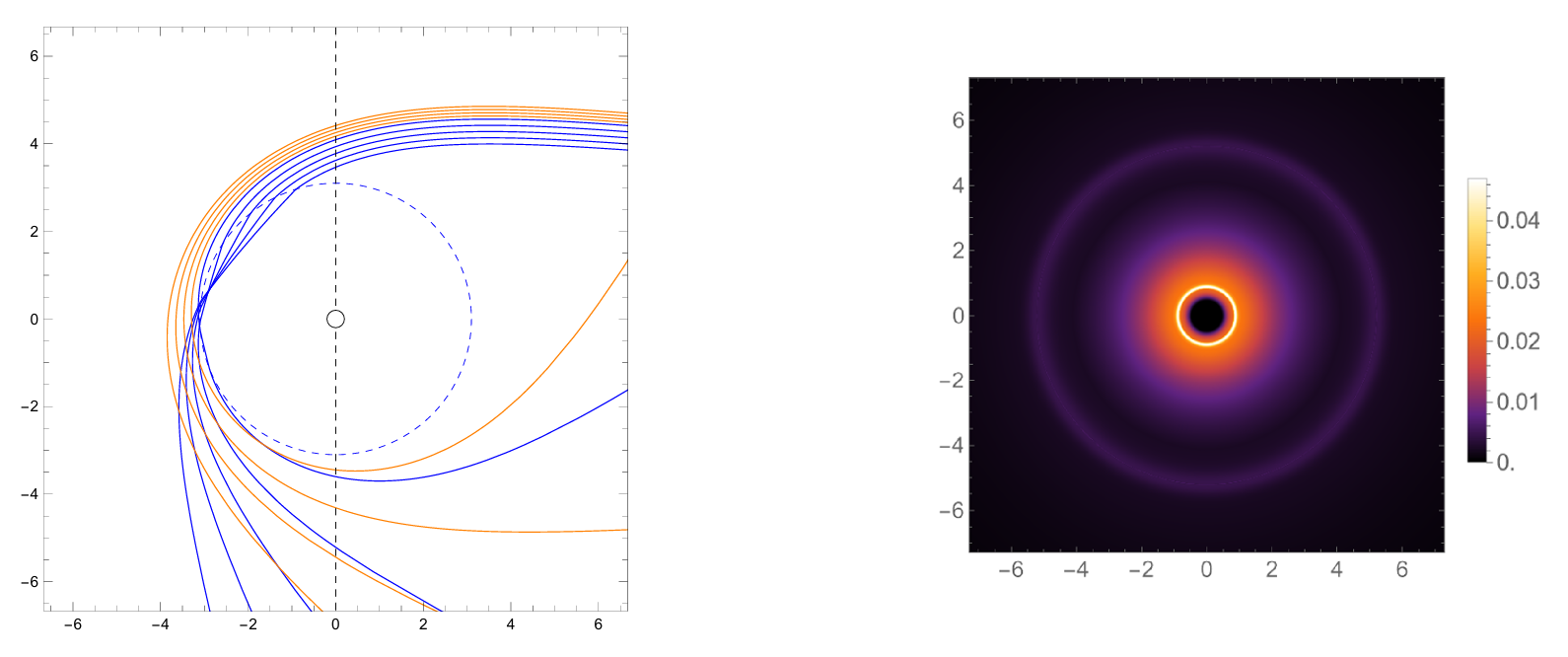}
    \caption{Left: ray trajectories for \(m_-=0.1\) and \(R_{\rm sh}=3.1\) that intersect the accretion disk twice. The orange trajectory does not cross the shell, while the blue trajectory does. The blue dashed circle marks the shell location, the black circle marks the inner Schwarzschild horizon, and the vertical black dashed line marks the accretion disk. Refraction at the shell is clear along the blue ray: rays with smaller impact parameter \(b\) reach a larger radius \(r\) at their second intersection with the disk. Right: black hole image corresponding to the intensity in Fig.~\ref{mminus0.1,mplus1,ms3.1Refractioneffect}(a). The bright central ring arises from the inner Schwarzschild photon sphere. A broader and fainter outer ring corresponds to the peak at large \(b\) in Fig.~\ref{mminus0.1,mplus1,ms3.1Refractioneffect}(a). It is not associated with an external photon sphere.}
    \label{mminus0.1,mplus1,ms3.1RefractioneffectTrajectory}
\end{figure}

Fig.~\ref{mminus0.1,mplus1,ms3.1Refractioneffect} shows the observed intensity \(I_{\rm obs}(b)\), the transfer function \(r(b)\), and the corresponding redshift factors \(g_{n}(b)\) for \(m_{-}=0.1\) and \(R_{\rm sh}=3.1\). 
Three notable features emerge.
First, panel~(c) shows cusps (non-differentiable points) on all three redshift curves, indicating that the redshift evaluated at the emission radius is not a smooth function.
Second, panel~(a) exhibits two peaks: one at relatively small \(b\), 
arising from the photon sphere of the inner Schwarzschild region,
and the other appearing near the critical impact parameter \(b_{c+}=3\sqrt{3}\) of the outer Schwarzschild region.
Although the outer photon sphere is excised by the shell,
since \(R_{\rm sh}=3.1>3\), 
so that there is no double photon sphere configuration in this case, 
the image still shows two photon rings.
The same behavior is visible in the right panel of Fig.~\ref{mminus0.1,mplus1,ms3.1RefractioneffectTrajectory}, which shows a brighter inner ring accompanied by a fainter outer ring.
Third, in panel~(b) near \(b=3\sqrt{3}\), the transfer functions \(r(b) \) for rays that intersect the accretion disk twice and three times (orange and red) form a pronounced V-shape: along the right branch \(r\) increases with \(b\), whereas along the left branch \(r\) increases as \(b\) decreases. 
Consequently, the nearby peak in panel~(a) becomes broader and takes on an approximately \(\Lambda\) shaped profile instead of the familiar sharp profile with a right shoulder drop that is typically associated with images of Schwarzschild black holes.

The reason for the redshift cusp can be understood as follows. The cusp corresponds to rays whose intersection radius \(r\) with the disk equals the shell radius \(R_{sh}\).
After introducing the thin shell, the four velocity of the static observers inside and outside is continuous but not smooth across the shell, as made explicit by Eq.~\eqref{eq:StaticTrans}. Accordingly, the redshift factor in Eq.~\eqref{eq:Staticgfactor} develops a cusp at the shell.
Therefore, whenever the emission point crosses the shell, the redshift curve develops a cusp. This happens for light rays that intersect the accretion disk once, twice, or three times.

The additional peak near \(b = 3\sqrt{3}\) is a consequence of the piecewise structure of the spacetime constructed with a shell.
The shell causes a discontinuity in geometric structures at \(R_{sh}\), while each side remains Schwarzschild.
For Schwarzschild spacetime, the gravitational field strengthens toward the black hole, giving rise to the photon sphere at \(r=3M\).
The strong-field region near the photon sphere can also cause multiple deflections of light.
In our setup, contributions from rays that intersect the disk four or more times are neglected, and trajectories with three intersections are usually taken to be dominated by photon sphere effects. 
When a thin shell is placed near the photon sphere radius, the exterior may no longer admit a true photon sphere. 
However, the strong-deflection structure in the vicinity of the photon sphere is preserved in the joined geometry.
As a result, a ring shaped brightness enhancement can appear near \(b_{c+}=3\sqrt{3}\), which produces an additional peak.
This also explains the double ring structure in the right panel of Fig.~\ref{mminus0.1,mplus1,ms3.1RefractioneffectTrajectory}, where the outer ring is not associated with an external photon sphere.

The V-shaped structure in Fig.~\ref{mminus0.1,mplus1,ms3.1Refractioneffect}(b) directly reflects the shell’s refraction effect. As derived in Sec.~\ref{sec:shell-dynamics-redshift}, rays refract when traversing the shell, altering their propagation direction.
The left panel of Fig.~\ref{mminus0.1,mplus1,ms3.1RefractioneffectTrajectory} shows rays that intersect the disk twice. The shell refracts rays with small \(b\) outward, which increases the radius \(r\) of the second intersection with the disk.
As a result, \(r(b)\) becomes nonmonotonic near \(b \approx 3\sqrt{3}\) and takes on a V-shape in the plot.
Similar V-shaped features of the transfer function \(r(b)\) have, to our knowledge, been reported mainly in spacetimes with naked singularities or wormholes~\cite{Luo:2023wru,Rueda:2025rlj,Tan:2025cte}.
In contrast, in standard black hole spacetimes without a refractive shell, \(r(b)\) is typically monotonic.
In our setup, we find that for suitable shell parameters, adding a refractive shell around a black hole can produce such a V-shaped structure in the transfer function.
The same behavior occurs for rays that intersect the disk three times.
This is the reason for the \(\Lambda\)-shaped peak in Fig.~\ref{mminus0.1,mplus1,ms3.1Refractioneffect}(a).

\begin{figure}[htbp]
    \centering
        \includegraphics[width= 1 \linewidth]{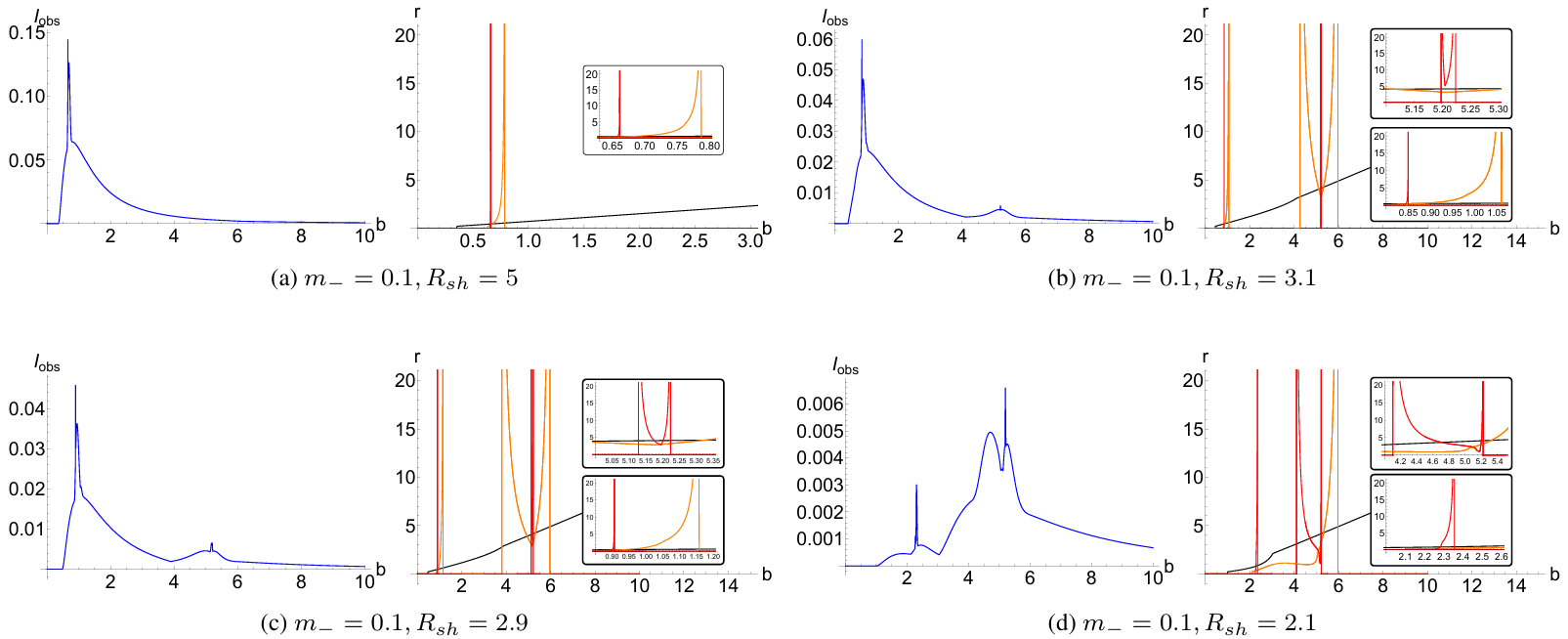}
\caption{Observed intensity profiles $I_{\rm obs}(b)$ and transfer functions $r(b)$ with $m_-=0.1$, 
shown for (a) $R_{\rm sh}=5$, (b) $R_{\rm sh}=3.1$, (c) $R_{\rm sh}=2.9$, and (d) $R_{\rm sh}=2.1$. 
In the right panels, black, orange, and red curves correspond to rays intersecting the disk once, twice, and three times, respectively.
Insets in the $r(b)$ panels provide zoomed-in views of the local features.}
\label{mminus0.1,mplus1Iob}
\end{figure}

\begin{figure}[htbp]
    \centering
        \includegraphics[width= 1 \linewidth]{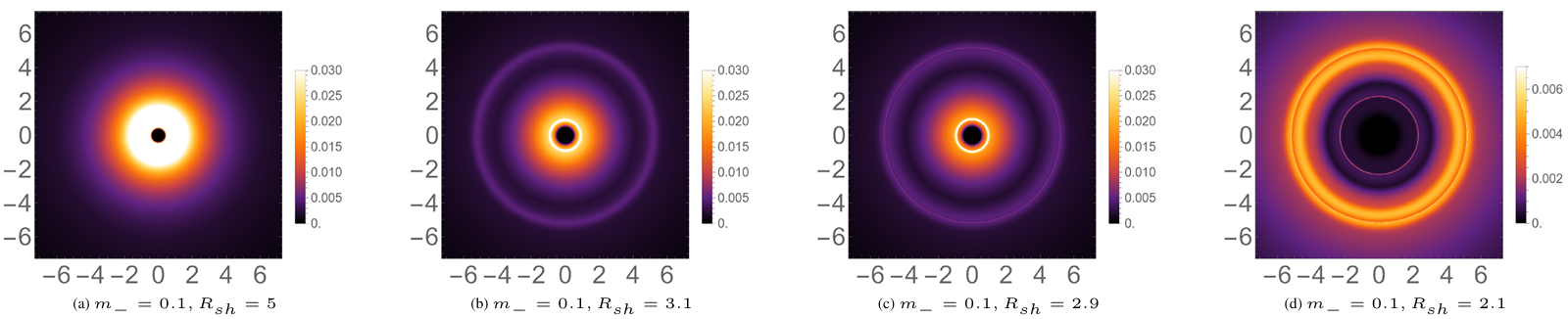}
\caption{Black hole images corresponding to the profiles in Fig.~\ref{mminus0.1,mplus1Iob}, 
for a static thin shell with fixed interior mass $m_-=0.1$. 
Panels (a)--(d) show the results for spherical shells at radii $R_{\rm sh}=5$, $3.1$, $2.9$, and $2.1$, respectively.
In panel (d), we use a different intensity scale in order to highlight the double photon ring structure.
}
\label{mminus0.1,mplus1BH}
\end{figure}

After clarifying the basic features of the images produced by a static shell, we examine how these features vary with the shell position.
Fig.~\ref{mminus0.1,mplus1Iob} shows the effect of decreasing \(R_{\rm sh}\) while keeping the inner mass fixed at \(m_{-}=0.1\). 
When \(R_{\rm sh}=5>3\), the shell lies far from the outer photon sphere and the intensity profile displays only the inner photon ring peak. 
When \(R_{\rm sh}=3.1\) (the case discussed above), 
additional peaks appear near \(b=b_{c+}=3\sqrt{3}\) that are not produced by an external photon sphere. 
When \(R_{\rm sh}=2.9\), the spacetime contains both inner and outer photon spheres, and the intensity still shows two peaks: 
the peak at smaller \(b\) originates from the inner photon sphere, and the peak at \(b=3\sqrt{3}\) corresponds to the outer photon sphere. In the right panel of Fig.~\ref{mminus0.1,mplus1Iob}(c), the orange and red curves again form a V shape. 
The right branch is controlled by the outer photon sphere, and the left branch is generated by refraction at the shell. 
Further decreasing \(R_{\rm sh}\) to \(2.1\) makes the double peak structure more pronounced.
As the shell moves away from the outer photon sphere, the V shape deforms: the right branch changes little, while the left branch shifts inward with \(R_{\rm sh}\). Concretely, the red curve continues to show a clear V shape, while the orange curve no longer exhibits an ideal V shape.

Fig.~\ref{mminus0.1,mplus1BH} presents the black hole images corresponding to Fig.~\ref{mminus0.1,mplus1Iob}. 
As $R_{\rm sh}$ decreases, the inner photon ring gradually moves outward and its brightness decreases, while the outer photon ring begins to appear.
The decrease in the brightness of the inner photon ring can be understood as follows: for a smaller shell radius $R_{\rm sh}$, photons emitted from the shell experience a larger redshift, so that the corresponding inner photon ring becomes dimmer on the observer screen.
The outward shift of the inner photon ring can be seen from the transformation Eq.~\eqref{eq:Statictrans}: for light emitted from the same inner photon sphere, a smaller shell radius $R_{\rm sh}$ leads to a larger corresponding impact parameter $b_+$ in the exterior region, and therefore to a larger photon-ring radius in the image.
As for the appearance of the outer photon ring, it is caused by the change in the photon sphere structure of the spacetime. As $R_{\rm sh}$ decreases, the structure changes from having only an inner photon sphere to having both inner and outer photon spheres, and an outer photon ring then appears in the image.
Specifically, in panel (a) the shell is still distant and only the inner photon ring is visible. As the shell moves closer to the outer photon sphere, panels (b) and (c) display a double ring structure, with the outer ring appearing dimmer due to its lower intensity. In panel (d), as the shell approaches the exterior Schwarzschild horizon, the double ring feature becomes increasingly pronounced.

\begin{figure}[htbp]
    \centering
        \includegraphics[width= 1 \linewidth]{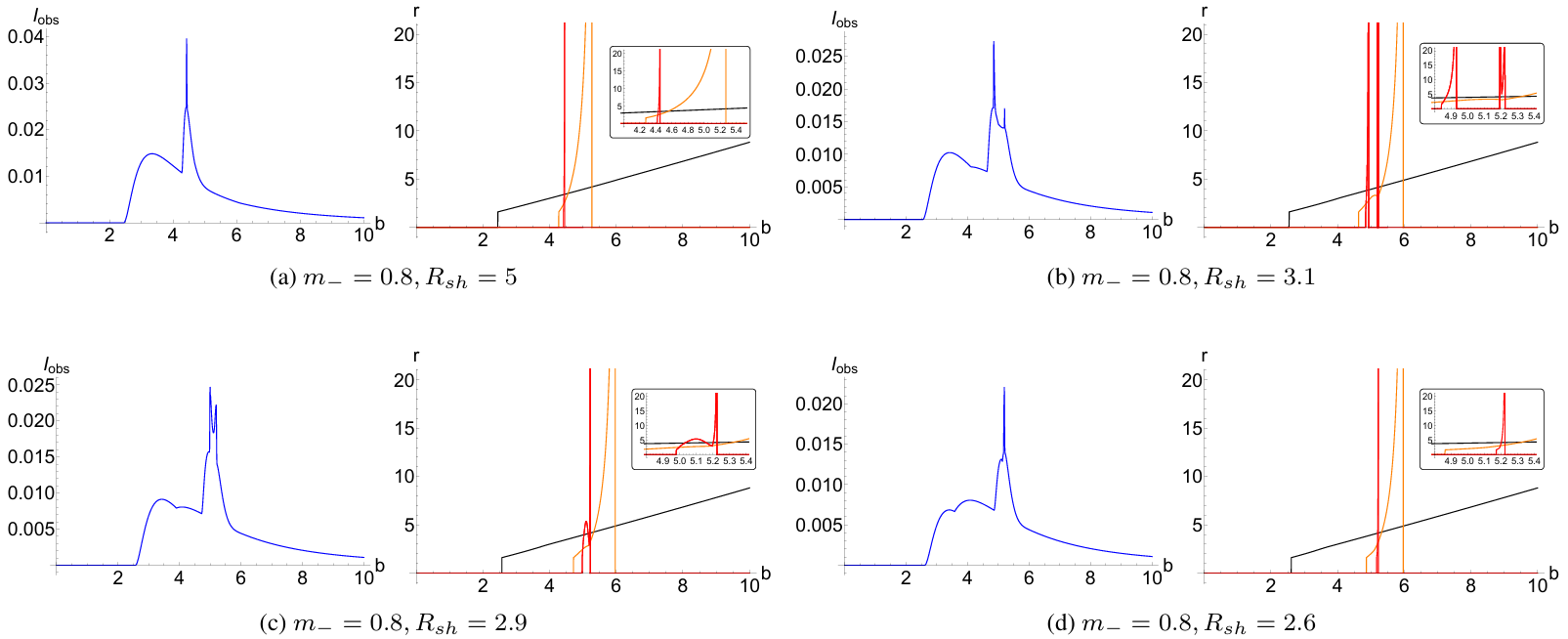}
\caption{Observed intensity profiles $I_{\rm obs}(b)$ and transfer functions $r(b)$ with $m_-=0.8$, 
shown for (a) $R_{\rm sh}=5$, (b) $R_{\rm sh}=3.1$, (c) $R_{\rm sh}=2.9$, and (d) $R_{\rm sh}=2.6$. 
In the right panels, black, orange, and red curves correspond to rays intersecting the disk once, twice, and three times, respectively.
Insets in the $r(b)$ panels provide zoomed-in views of the local features.}
\label{mminus0.8,mplus1Iob}
\end{figure}

\begin{figure}[htbp]
    \centering
        \includegraphics[width= 1 \linewidth]{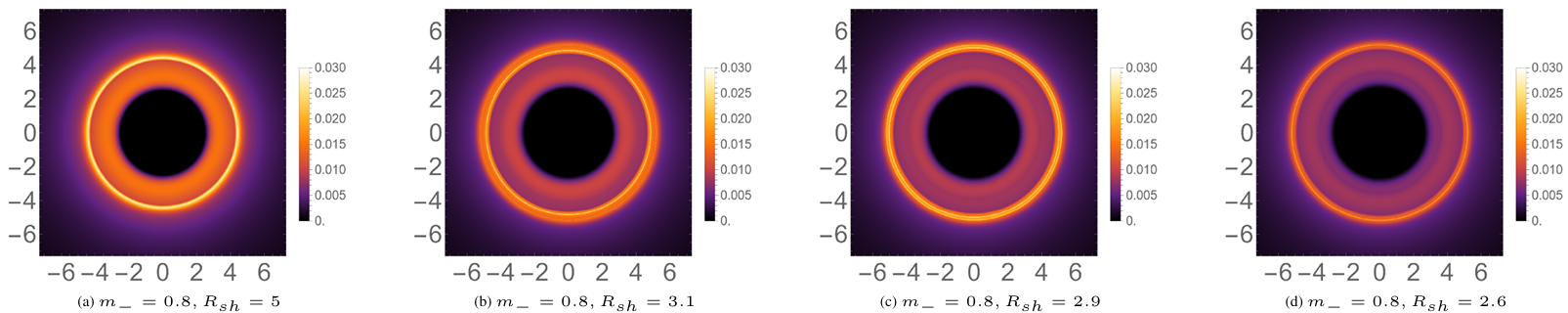}
\caption{Black hole images corresponding to the profiles in Fig.~\ref{mminus0.8,mplus1Iob}, 
for a static thin shell with fixed interior mass $m_-=0.8$. 
Panels (a)--(d) show the results for spherical shells at radii $R_{\rm sh}=5$, $3.1$, $2.9$, and $2.6$, respectively.}
\label{mminus0.8,mplus1BH}
\end{figure}

Fig.~\ref{mminus0.8,mplus1Iob} examines the case with a larger inner mass \(m_{-}=0.8\), again decreasing \(R_{\rm sh}\). 
For \(R_{\rm sh}=5\), \(3.1\), and \(2.9\), the overall behavior is similar to that in Fig.~\ref{mminus0.1,mplus1Iob}. A small difference appears for \(R_{\rm sh}=2.9\): in the right panel of Fig.~\ref{mminus0.8,mplus1Iob}(c), the red curve no longer shows a pronounced V shape. This is because the mass difference between the interior and exterior is smaller, so the refraction effect is weaker and the V-shaped feature becomes less prominent.
Finally, for \(R_{\rm sh}=2.6\), the spacetime still contains two photon spheres, but the intensity profile shows only a single peak. According to Eq.~\eqref{eq:Statictrans}, the inner critical impact parameter \(b_{c-}=3\sqrt{3}\,m_{-}\) is mapped by the shell transformation to an exterior value \(b_{+}\) that exceeds the exterior Schwarzschild critical impact parameter \(b_{c+}=3\sqrt{3}\). In this situation, rays with \(b\simeq b_{c-}\) will be absorbed by the shell and therefore do not contribute to the observed image. From the observer’s point of view, only the peak associated with the outer photon sphere remains visible.

Fig.~\ref{mminus0.8,mplus1BH} presents the black hole images corresponding to Fig.~\ref{mminus0.8,mplus1Iob}. 
The overall trend follows that of Fig.~\ref{mminus0.1,mplus1BH}.
The total intensity decreases, and the rings shift outward. 
In panel (a), only the inner photon ring is present. In panel (b), a double ring structure appears and the inner ring remains brighter. In panel (c), the two rings have comparable brightness. In panel (d), the inner photon ring disappears and only the outer photon ring remains.

In summary, a static thin shell spacetime exhibits several distinctive observational features.
The shell affects the gravitational redshift, and it induces refraction at the interface. More importantly, as \(R_{\rm sh}\) decreases, the photon sphere configuration changes from an inner photon sphere to a double photon sphere configuration and finally to an outer photon sphere. However, the observed intensity profile does not follow these changes one to one and instead responds in a continuous and gradual manner.
Specifically, before a double photon sphere forms (for example, \(R_{\rm sh}=3.1\) with \(m_{-}=0.1\) or \(0.8\)), the intensity profile can already develop two peaks. 
Conversely, even when two photon spheres are present (for example, \(m_{-}=0.8\) and \(R_{\rm sh}=2.6\)), the intensity profile may show only a single peak. 
Thus the spacetime geometry and the observed intensity profile are not in a one-to-one correspondence, and their relation is governed by the combined effects introduced by the shell.

\section{Dynamic Shell Imaging}\label{sec:DynamicImaging}

For Schwarzschild spacetimes joined by a static thin shell, the shell introduces several key features in the image. 
First, the redshift factor is not smooth at the shell, 
and this produces visible cusps in the intensity profile. 
Second, there is refraction at the shell: when the shell approaches the outer photon sphere, rays in the photon sphere neighborhood refract at the interface, causing the transfer function to develop a characteristic V-shaped profile.
Finally, regarding the correspondence between photon spheres and photon rings, as the shell position varies the geometry may admit two photon spheres, yet the image may not exhibit a double photon ring, and the converse can also occur.
Building on these static features, we now turn to images produced by a collapsing shell. In this dynamical setting, we also take light travel time delays into account.
We examine whether these features are still present when the shell moves and identify new phenomena caused by the shell motion.

We begin with the shell dynamics. As shown in Sec.~\ref{sec:thin-shell}, whether the shell reaches infinity is controlled by the parameter \(e\) as Eq.~\eqref{eq:Rdot-simple}. In our calculations we take \(e=1\) by default, so the shell can reach infinity with vanishing asymptotic velocity.
We make this choice for two reasons. For \(e>1\), the qualitative features of the image do not change. For \(e<1\), the shell reaches only a finite maximum radius \(R_{\max}\). If \(R_{\max}\) is not close to a photon sphere, the resulting images are typically similar to the \(e=1\) case.
We have also made available online videos of time-dependent black hole images for various values of \(e\) \footnote{https://github.com/Xia-YuanLiu/photon-ring-thin-shell-videos}. 
To extract dynamical shell features in a simple setting, we focus on \(e=1\) in what follows.

From Eq.~\eqref{eq:dRdt} one sees that the evolution of the shell radius differs when measured with the interior and exterior time coordinates \(t_{-}\) and \(t_{+}\). We impose the common initial condition \(R=10\) at \(t_{-}=t_{+}=0\). When a photon crosses the shell, its four momentum components transform according to Eq.~\eqref{eq:kTrans-restate}, and the time coordinates are switched \((t_{-}\leftrightarrow t_{+})\). The remaining setup follows the static shell case: the accretion disk extends from the inner Schwarzschild horizon out to infinity, with emissivity given by Eq.~\eqref{accretiondisk}. The observer is placed at \((t_{o},\, r_{o}=50,\, \phi_{o}=0)\), and we fix \(\theta=\pi/2\) and \(m_{+}=1\).

\begin{figure}[htbp]
    \centering
        \includegraphics[width= 1 \linewidth]{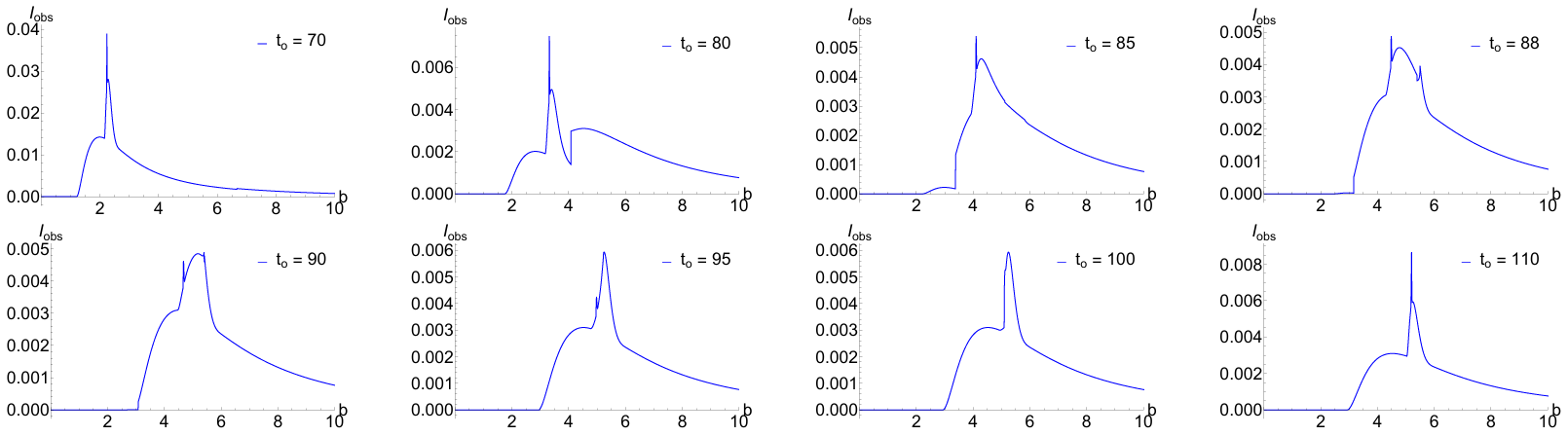}
\caption{Observed intensity \(I_{\rm obs}\) at multiple \(t_o\) for \(m_-=0.3\), \(m_{\mathrm{sh}}=0.7\).}
\label{mMinus0.3ms0.7Rsh10Iob}
\end{figure}

\begin{figure}[htbp]
    \centering
        \includegraphics[width= 1 \linewidth]{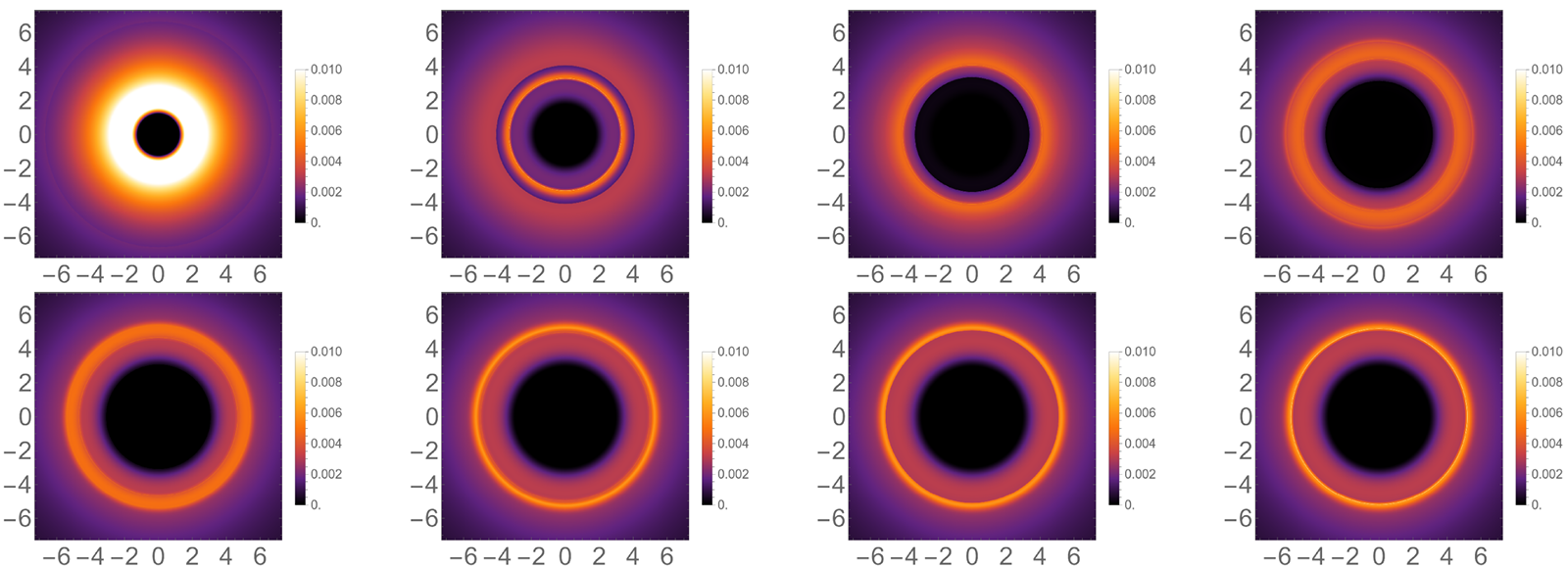}
\caption{Black hole image at multiple \(t_o\) for \(m_-=0.3\), \(m_{\mathrm{sh}}=0.7\).}
\label{mMinus0.3ms0.7Rsh10BH}
\end{figure}

Fig.~\ref{mMinus0.3ms0.7Rsh10Iob} shows the intensity profile at different observation times \(t_{o}\) for \(m_{-}=0.3\) and \(m_{\mathrm{sh}}=0.7\). Because of light travel time delays, at fixed \(t_{o}\) rays with different impact parameters \(b\) are emitted at different times \(t_{e}\).
Each image at a given \(t_{o}\) is therefore a superposition of signals emitted over an interval \([t_{e1},t_{e2}]\). 
As the observation time $t_{o}$ increases, the peak moves to larger $b$.
From $t_{o}=70$ to $t_{o}=80$, the overall intensity drops rapidly and then becomes almost steady.
This trend is similar to the static shell case where the shell radius is gradually decreased, and the physical reason is also similar. 
As the shell radius 
$r$ becomes smaller, the redshift becomes stronger and the observed intensity decreases.
At the same time, light associated with the inner photon sphere is mapped to larger $b_+$, so the peak shifts to the right.
The only difference is that in the static shell case the shell position is changed by hand, whereas here the change is caused by the collapse of the shell as time increases. 
The corresponding images in Fig.~\ref{mMinus0.3ms0.7Rsh10BH} show the same trend: at \(t_{o}=70\) the image is very bright, and as time increases the total intensity becomes nearly steady, while the photon ring shifts outward from an inner ring toward the outer ring.
This behavior tracks the continuing inward collapse of the shell.
At earlier times, when the shell radius \(R\) is larger, the inner ring is visible, and once the shell passes inside the exterior Schwarzschild horizon the image evolves into the outer photon ring.

In addition to this overall trend, we also find two important features in the images produced by the dynamical shell.
First, step-like discontinuities appear in the intensity profile, most clearly at \(t_{o}=80\).
Second, we see double peaks at \(t_{o}\approx 88, 90, 95\).
During the collapse, the photon sphere structure changes from an inner photon sphere, to a configuration with two photon spheres, and finally to an outer photon sphere. One might therefore expect to see a double photon ring in the image at some intermediate times.
However, the double peaks observed here are not the double photon ring structure we would expect.
We will explain the reasons for these two points in detail below.

\begin{figure}[htbp]
    \centering
        \includegraphics[width= 1 \linewidth]{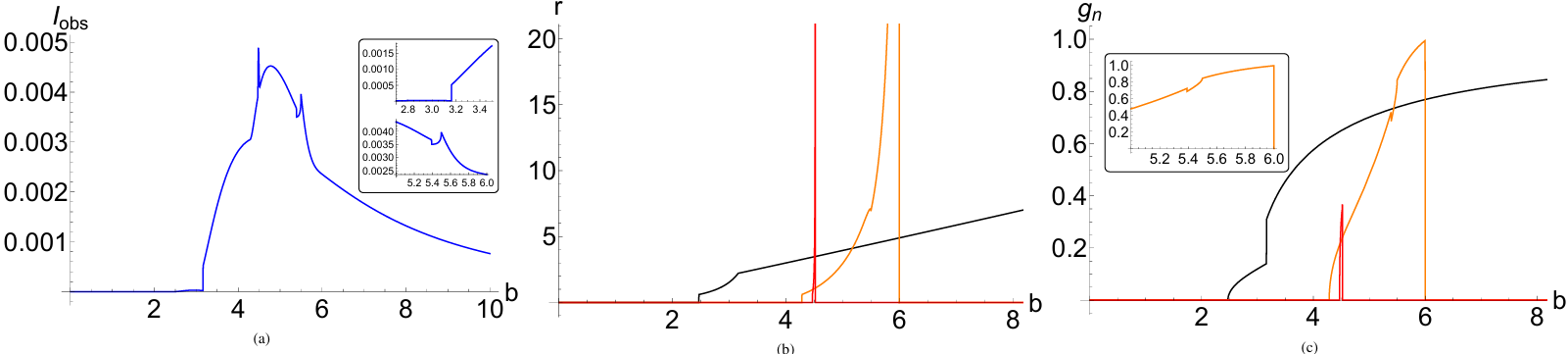}
    \caption{Parameters \(m_{-}=0.3\), \(m_{\mathrm{sh}}=0.7\), and observation time \(t_{o}=88\).
(a) Observed specific intensity profile \(I_{\rm obs}(b)\).
(b) Transfer function \(r(b)\).
(c) Corresponding redshift factors \(g_{n}(b)\).
In panels (b) and (c), black, orange, and red curves denote rays that intersect the disk once, twice, and three times, respectively.}
    \label{NoDoublePeakPhenomenon}
\end{figure}

\begin{figure}[htbp]
    \centering
        \includegraphics[width= 1 \linewidth]{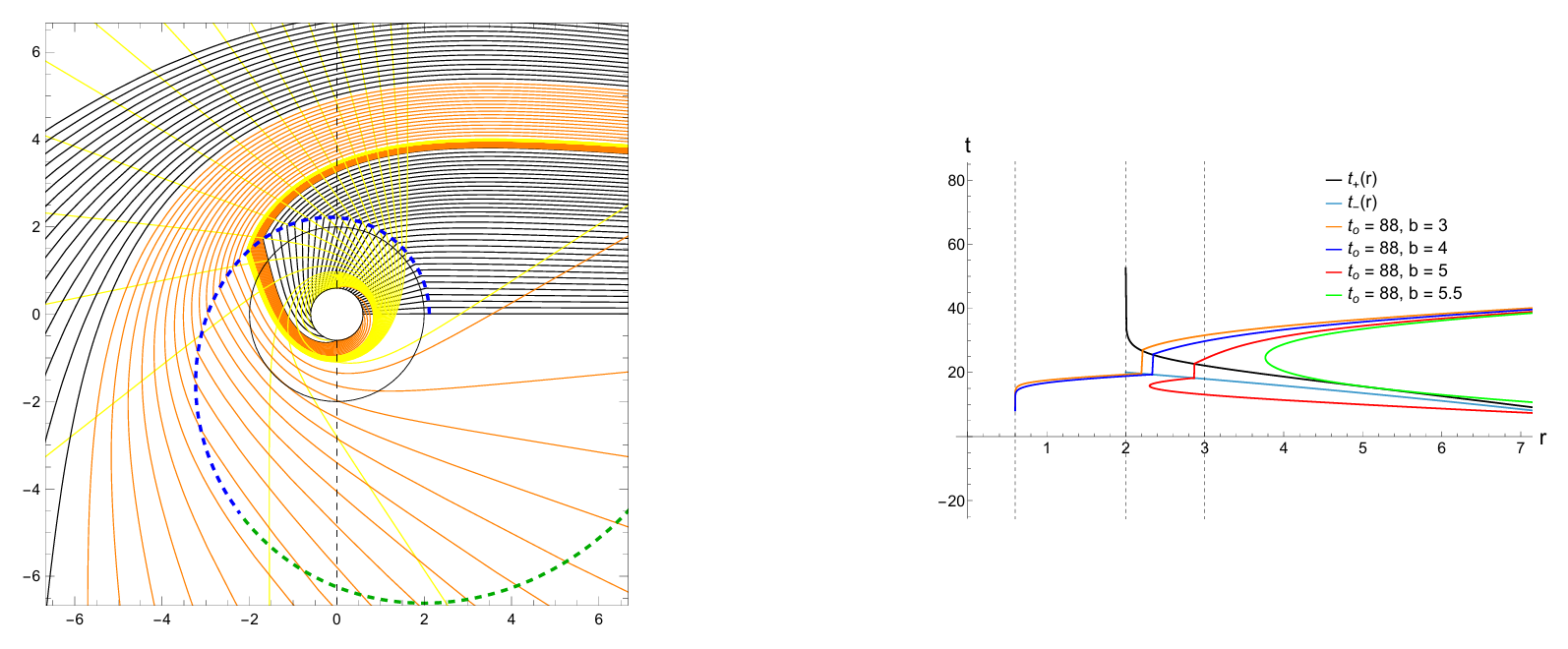}
    \caption{Left: ray trajectories for \(m_{-}=0.3\), \(m_{\rm sh}=0.7\), \(t_{o}=88\).
Black, orange, and yellow curves denote rays intersecting the disk once, twice, and three or more times, respectively.
The small and large black circles mark the inner and outer Schwarzschild horizons.
The vertical black dashed line marks the disk.
The blue dashed curve traces exit points where rays cross the shell from the interior to the exterior.
The green dashed curve traces entry points where rays cross the shell from the exterior to the interior.
Right: worldlines for different exterior impact parameters \(b_+ \).
The black curve shows the shell in the exterior time coordinate, and the light blue curve shows the shell in the interior time coordinate.
The orange, blue, and red curves represent rays whose exterior impact parameters are \(b_{+}=3\), \(b_{+}=4\), and \(b_{+}=5\), respectively.
The green curve corresponds to the critical case where the ray just touches the shell worldline.
}
    \label{NoDoublePeakPhenomenon2}
\end{figure}

\begin{figure}[htbp]
    \centering
        \includegraphics[width= 1 \linewidth]{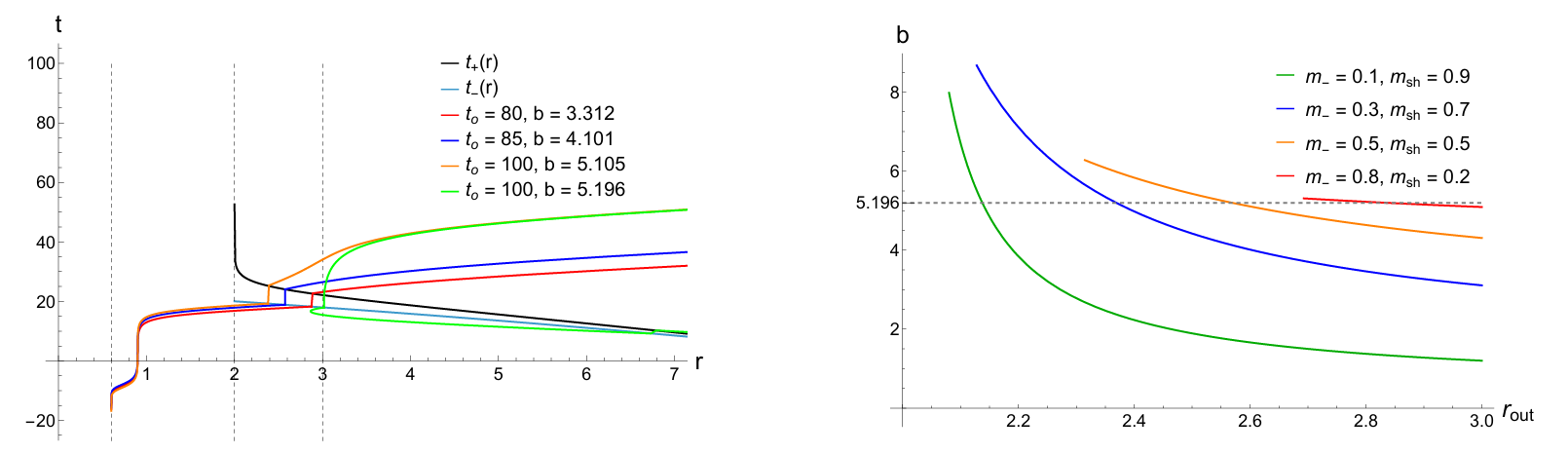}
    \caption{Left: worldlines for different impact parameters \(b\) with \(m_{-}=0.3\) and \(m_{\rm sh}=0.7\).
The black and light blue curves are the shell worldlines in the exterior and interior time coordinates.
The orange, blue, and red curves represent rays from the inner photon sphere that exit the shell at different radii \(r_{\rm out}\), characterized by
exterior impact parameters \(b_{+}=3.312,\,4.101,\,5.105\) and observation times \(t_{o}=80,\,85,\,100\), respectively.
The green curve represents a ray from the outer photon sphere observed at \(t_{o}=100\).
Rays associated with the inner photon sphere can be observed together with rays associated with the outer photon sphere only when they exit at sufficiently small \(r_{\rm out}\).
Right: for different interior masses \(m_{-}\) (with \(e=1\) fixed), the exterior impact parameter \(b_{+}\) as a function of the exit radius \(r_{\rm out}\).
}
    \label{NoDoublePeakExplanation}
\end{figure}

For definiteness we take \(t_{o}=88\). Fig.~\ref{NoDoublePeakPhenomenon} for \(m_{-}=0.3,\; m_{\mathrm{sh}}=0.7,\; t_{o}=88\) presents three panels: the intensity profile, the transfer function \(r(b)\), and the redshift factors \(g_{n}(b)\).
We first discuss the step-like features.
In Fig.~\ref{NoDoublePeakPhenomenon}(a) the intensity shows two steps, one near \(b\simeq3.16\) and the other near \(b\simeq5.4\). These match one-to-one the jumps in the black and orange curves of Fig.~\ref{NoDoublePeakPhenomenon}(c) and arise from discontinuities of the redshift factor. 
The same reasoning as in the static case applies: the shell makes the four velocity of the static observer inside and outside different. 
When the shell is at rest, the four velocity of the static observer is continuous across the shell,
whereas once the shell moves it becomes discontinuous. 
In addition, in the dynamical case the redshift depends not only on the emission radius, but also on how many times the ray passes through the shell, on the radius at which it crosses the shell, and on the shell velocity.

Fig.~\ref{NoDoublePeakPhenomenon2} (left) shows ray trajectories in the dynamical shell spacetime. 
Black, orange, and yellow curves denote rays that intersect the disk once, twice, and three or more times, respectively. The blue curve is the locus of exit points where rays cross the shell from the interior to the exterior region. The dark green curve is the locus of entry points where rays cross the shell from the exterior into the interior. The vertical black line marks the disk.
The right panel shows worldlines at \(t_{o}=88\) for different \(b\). 
By matching the locations of these discontinuities, we see that the jump of the black curve in Fig.~\ref{NoDoublePeakPhenomenon}(c) at \(b\simeq 3.16\) corresponds to rays whose first intersection with the accretion disk occurs at the same radius where they emerge from the shell 
(the intersection point between the blue dashed curve and the black vertical dashed line in the left panel of Fig.~\ref{NoDoublePeakPhenomenon2}). 
For \(b\) slightly larger than \(3.16\), the observed rays do not pass through the shell, whereas for \(b\) slightly smaller than \(3.16\), they pass through the shell once, which produces the jump in the redshift factor.
Similarly, the jump of the orange curve near \(b \simeq 5.4\) in Fig.~\ref{NoDoublePeakPhenomenon}(c) corresponds to rays whose second intersection with the accretion disk coincides with the radius where they pass through the shell, namely the point where the dark green dashed curve intersects the black vertical dashed line.

Next we examine the double peak phenomenon. 
Generally, we regard a peak as arising from a photon sphere if rays with \(b\) in some interval around the peak intersect the disk at least three times.
Fig.~\ref{NoDoublePeakPhenomenon}(b) shows that only one peak satisfies this criterion, with the true peak located near the red curve at \(b\simeq 4.5\). 
Therefore, the peak near \(b\simeq 5.5\) in Fig.~\ref{NoDoublePeakPhenomenon}(a) does not correspond to a photon sphere.
In Fig.~\ref{NoDoublePeakPhenomenon2} (right), we plot the photon worldlines for different values of \(b\). The green worldline corresponds to the peak discussed here, while the black and light blue curves show the shell position as a function of the exterior and interior times, respectively. We see that the green curve is just tangent to the black one, which means that the corresponding rays are not affected by the shell.
In the ray trajectory diagram in Fig.~\ref{NoDoublePeakPhenomenon2} (left), this case corresponds to the orange trajectory, which is tangent to both the blue dashed curve and the dark green dashed curve. The tangency point is exactly the intersection of these two dashed lines.
For \(b<5.5\) refraction at the shell modifies the trajectory, whereas for \(b>5.5\) the ray is unaffected by the shell. 
We therefore attribute this peak primarily to refraction at the shell.

Why, then, does a double photon ring not appear?
To resolve a peak produced by the outer photon sphere, several conditions must be met. The shell must first arrive in the vicinity of \(r=3\). 
A photon emitted from the disk then orbits multiple times near the photon sphere, intersecting the disk at least three times before reaching the observer.
This takes a finite propagation time, during which the shell continues to collapse. 
During this time window, there are two possibilities. If the shell has already fallen inside the exterior Schwarzschild horizon \((r=2)\), interior emission cannot escape and no double photon ring forms. If the shell is still outside the horizon during this interval, interior light can emerge and be detected.

Fig.~\ref{NoDoublePeakExplanation} makes this competition explicit. In the left panel, the red, blue, and orange curves are worldlines of rays associated with the inner photon sphere, and the green curve corresponds to the outer photon sphere. 
The right panel plots, for various interior masses \(m_{-}\), the exterior impact parameter \(b_{+}\) of rays from the inner photon sphere when they emerge from the shell at \(r_{\text{out}}\).
Although the impact parameter of the inner photon sphere is \(b_{c-}=3\sqrt{3}\,m_{-}\), when the rays cross the shell their energy is rescaled, and the corresponding \(b_{+}\) is shifted.
This rescaling depends on the shell velocity at the crossing, which in turn depends on the crossing radius, so \(b_{+}\) is ultimately a function of \(r_{\text{out}}\).
As the number of windings around the outer photon sphere increases, the travel time of the corresponding rays also grows. From the left panel of Fig.~\ref{NoDoublePeakExplanation}, we see that we only start to receive light from the outer photon sphere at observation times \(t_o \approx 100\). For smaller \(t_o\), the shell has not yet collapsed inside the outer photon sphere, so no light from that region can escape. 
As \(t_o\) increases, the exit radius \(r_{\text{out}}\) of rays associated with the inner photon sphere decreases, and the corresponding exterior impact parameter \(b_{+}\) increases toward the critical value \(b_{c+}=3\sqrt{3}\). 
Consequently, near \(b_{c+}\) the contributions from inner photon sphere rays and outer photon sphere rays merge, and the observed peak is the superposition of both.

The right panel of Fig.~\ref{NoDoublePeakExplanation} also shows that, as \(m_{-}\) increases (still taking \(e=1\)), the left endpoint of the curve shifts rightward.
For example, with \(m_{-}=0.8\) and \(m_{\mathrm{sh}}=0.2\), rays associated with the inner photon sphere cannot exit at sufficiently small \(r_{\text{out}}\), because after passing through the shell their exterior impact parameter exceeds the exterior critical value \(b_{c+}\).
This is consistent with the static shell analysis. 
In other words, the double photon ring has already vanished in this case before the shell reaches the inner photon sphere radius \(r=3m_{-}=2.4\).

\begin{figure}[htbp]
    \centering
        \includegraphics[width= 1 \linewidth]{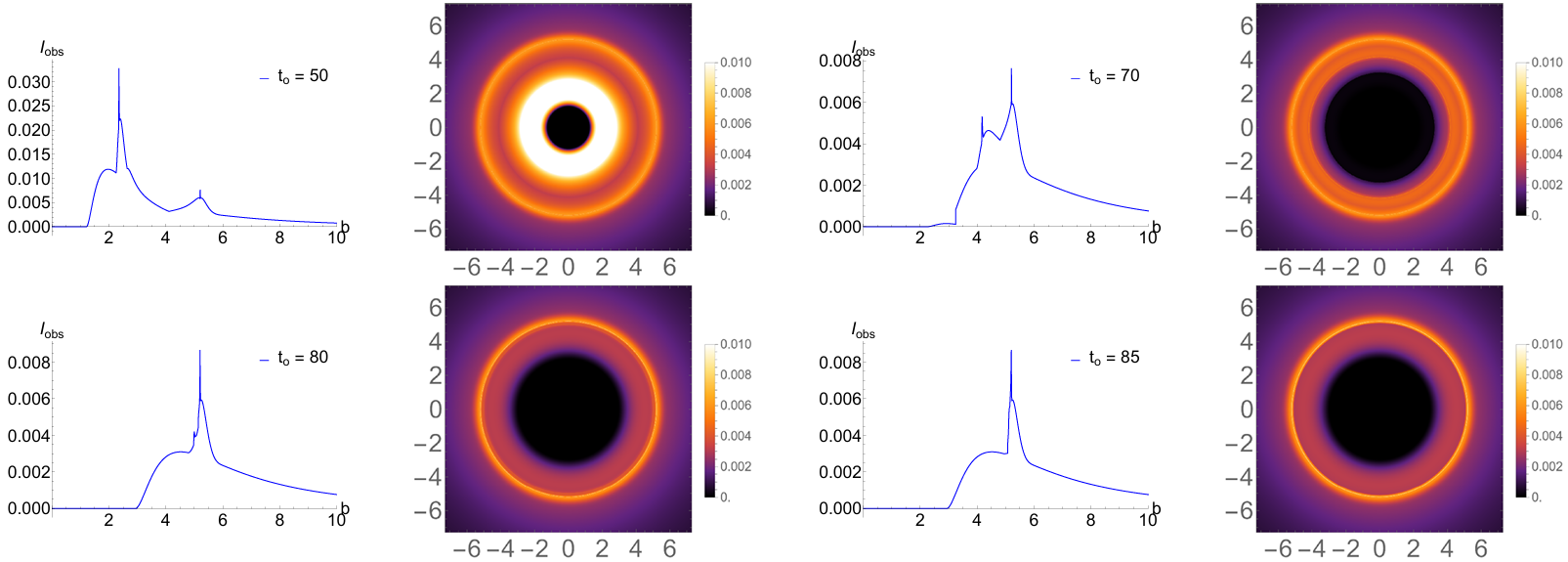}
\caption{Observed intensity \(I_{\rm obs}\) and black hole image at multiple \(t_o\) for \(m_-=0.3\), \(m_{\mathrm{sh}}=0.937\).}
\label{mMinus0.3ms0.937Rsh3.1}
\end{figure}

Motivated by the preceding observations, we seek a configuration in which rays associated with the outer photon sphere can undergo multiple windings and form the outer ring peak, while rays associated with the inner photon sphere can still exit at a relatively large \(r_{\text{out}}\) so that a genuine double photon ring is resolved. 
We take \(m_{-}=0.3\) and \(m_{\mathrm{sh}}=0.937\). 
In this case \(e<1\) and a maximum radius \(R_{\max}=3.1\) exists. 
The shell is kept static at \(r=3.1\) for \(t<0\) and begins to collapse inward from \(r=3.1\) at \(t_{-}=t_{+}=0\). 
Fig.~\ref{mMinus0.3ms0.937Rsh3.1} shows the intensity profiles and the corresponding images at different \(t_{o}\). 
At \(t_{o}=50\) the image is overall bright and a double photon ring appears, which is consistent with the static shell results since this time is early. 
At \(t_{o}=70\) the total intensity decreases and begins to stabilize, and the two peaks start to evolve, forming a bright band. 
The left peak, originating from the inner photon sphere, shifts to larger \(b\). This shift occurs because the corresponding rays cross the moving shell. As shown in Fig.~\ref{NoDoublePeakExplanation} (right), a smaller exit radius \(r_{\text{out}}\) maps to a larger exterior \(b_{+}\), and as a result the peak moves rightward.
The right peak, sourced by the outer photon sphere, remains nearly fixed. 
At \(t_{o}=80\) the left peak becomes barely visible. By \(t_{o}=85\) the shell has fully collapsed and the image approaches the standard Schwarzschild case. 
In this configuration, the geometry evolves from an inner photon sphere to a configuration with two photon spheres, and finally to an outer photon sphere.
From the image perspective, the appearance evolves from a double photon ring to a single outer photon ring.

\section{Conclusion}\label{Conclusion}

We examine the observational features that may arise in images of thin shell black holes constructed via the Israel junction conditions. 
To this end, we derive the dynamics of a pressureless shell, a transmission law for null geodesics across the shell, and a product formula for the total redshift when a light ray crosses the shell multiple times.
Building on these tools, we perform ray tracing of a geometrically and optically thin accretion disk and obtain images of Schwarzschild black hole with both static and collapsing shells.

In the static case, three characteristic features emerge. (i) Because the four velocity of static observers inside and outside the shell does not match smoothly, the photon energy changes at the crossing, producing a redshift cusp at the shell. 
(ii) The shell induces refraction. When the shell lies near the photon sphere radius, this refraction becomes especially strong and causes the transfer function \(r(b)\) to develop a V-shaped structure, which in turn produces a \(\Lambda\)-shaped photon ring peak in the observed intensity profile \(I_{\rm obs}(b)\).
(iii) The photon sphere structure and the photon ring peaks are no longer in one-to-one correspondence: an outer peak with a ring shape may appear even without an exterior photon sphere, while two photon spheres do not necessarily produce two photon ring peaks.

In the dynamical case, delays due to light travel and shell motion introduce time-dependent effects.
The intensity profile develops step-like discontinuities due to jumps in the redshift factor when rays cross the shell.
More importantly, although the geometry evolves from an inner photon sphere to a double and then to an outer photon sphere, the image typically does not display a clear double photon ring. When a double peak does appear, one peak is contributed by the inner photon sphere, while the other is produced by refraction at the shell rather than by a genuine second photon sphere.
We attribute this to propagation delays and identify a finely tuned case with \(e<1\) in which the shell remains near \(r \simeq 3\), yielding a genuine double photon ring that is resolvable for a short time window.

Compared to images of other compact objects, thin shell Schwarzschild black holes show many unique features. This is because the spacetime metric is not smooth everywhere in this case, and the spherical shell provides a very special structure. By contrast, spacetimes such as the Schwarzschild black hole
\cite{Gralla:2019xty}
or compact fluid stars without a thin shell
\cite{Rosa:2023hfm}
are smooth everywhere, so features like a cusp in the redshift factor and a breakdown of the correspondence between the photon sphere and the photon ring do not appear. For a thin shell Schwarzschild wormhole
\cite{Peng:2021osd},
a similar structure also appears at the throat, so its transfer function can show a V-shape and multiple photon rings. However, the overall geometry of a wormhole is still very different from what we discuss here, especially its photon sphere structure, so even photons with the same conserved quantities can follow different trajectories, which leads to different images overall.

In this work, our main goal was to highlight how the photon sphere structure of the junction spacetime is reflected in the observed images. For this reason, we adopted an idealized accretion disk model that always extends from the inner Schwarzschild horizon to infinity and did not include any backreaction of the shell on the disk itself. With this setup, the main signatures we find are largely insensitive to the radial emissivity profile of the disk and are governed primarily by the junction geometry.

A natural next step is to relax these idealizations by including the interaction between the shell and the accretion flow, considering rotating backgrounds, and allowing for shells of finite thickness or with nonzero pressure governed by an explicit equation of state. In particular, it would be interesting to study Kerr–to–Kerr matching as a rotating generalization of the setup considered here.
Including plasma dispersion and polarized radiative transfer would allow frequency–dependent and polarization–resolved predictions, and would provide a more stringent test of whether the imaging features identified here persist once these additional effects are taken into account. Overall, we highlight observational features characteristic of Israel junction spacetimes, laying the groundwork for their future observational testing.

\section*{Acknowledgement}

This work
was supported in part by the National Natural Science Foundation of China with grants No.12475063, No.12075232, No.12247103 and National Natural Science Foundation of China key project under Grant No. 12535002.

\bibliographystyle{unsrt}
\bibliography{main}

\end{document}